\documentclass[journal]{IEEEtran}
\usepackage{cite}
\usepackage[linesnumbered,ruled,vlined]{algorithm2e}
\usepackage{graphicx}
\usepackage{textcomp}
\usepackage{xcolor}
\usepackage{comment}
\usepackage{multicol}
\usepackage{multirow}
\usepackage{soul}
\newtheorem{example}{Example}
\usepackage{balance}

\usepackage{braket}
\usepackage{mathtools}
\usepackage{amsmath,amssymb,amsfonts}
\usepackage{algorithmic}
\usepackage{graphicx}
\usepackage{textcomp}
\usepackage{graphicx}
\usepackage{caption}
\usepackage{subcaption}

\usepackage{xcolor}
\def\BibTeX{{\rm B\kern-.05em{\sc i\kern-.025em b}\kern-.08em
    T\kern-.1667em\lower.7ex\hbox{E}\kern-.125emX}}
\newtheorem{definition}{Definition}

\usepackage{amsmath,amssymb,amsfonts}
\usepackage{algorithmic}
\usepackage{graphicx}
\usepackage{textcomp}
\def\BibTeX{{\rm B\kern-.05em{\sc i\kern-.025em b}\kern-.08em
    T\kern-.1667em\lower.7ex\hbox{E}\kern-.125emX}}
\begin{document}
%\history{Date of publication xxxx 00, 0000, date of current version xxxx 00, 0000.}
%\doi{10.1109/TQE.2020.DOI}

\title{Pauli Error Propagation-Based Gate Rescheduling for Quantum Circuit Error Mitigation}

\author{\IEEEauthorblockN{Vedika Saravanan, Samah Mohamed Saeed\\}
\IEEEauthorblockA{City College of New York, 
City University of New York, 
New York, USA\\
vsarava000@citymail.cuny.edu, ssaeed@ccny.cuny.edu}
%\and
%\IEEEauthorblockN{Samah Mohamed Saeed}
%\IEEEauthorblockA{City College of New York, 
%City University of New York\\
%New York, USA\\
%ssaeed@ccny.cuny.edu}
}

\maketitle
\begin{abstract}
Noisy Intermediate-Scale Quantum (NISQ) algorithms, which run on noisy quantum computers should be carefully designed to boost the output state fidelity. While several compilation approaches have been proposed to minimize circuit errors, they often omit the detailed circuit structure information that does not affect the circuit depth or the gate count. In the presence of spatial variation in the error rate of the quantum gates, adjusting the circuit structure can play a major role in mitigating errors. In this paper, we exploit the freedom of gate reordering based on the commutation rules to show the impact of gate error propagation paths on the output state fidelity of the quantum circuit, propose advanced predictive techniques to project the success rate of the circuit, and develop a new compilation phase post-quantum circuit mapping to improve its reliability. Our proposed approaches have been validated using a variety of quantum circuits with different success metrics, which are executed on IBM quantum computers. Our results show that rescheduling quantum gates based on their error propagation paths can significantly improve the fidelity of the quantum circuit in the presence of variable gate error rates.
\end{abstract}


%\titlepgskip=-15pt

\maketitle

%\balance

\section{Introduction}
%\todo{you need to read at least 1 paper a day to improve your writing. The first paragraph can be something you say in a presentation but not in scientific paper ex (when it comes to, don't say researchers started to show interest... be specific. How is it serving our problem. Try to avoid that something will be explained later unless it is necessarily. Here you don't need to classify noise  to carry the discussion. you don't execute a circuit on a software but on a computer. the simple model is based on errors computed during calibration and not success rate}

%Noisy Intermediate-Scale Quantum (NISQ) computers are anticipated to speed up the computational power of classical computers for certain classes of problems~\cite{Preskill_2018}. {\color{black} However, they suffer from a small number of noisy qubits in the range of tens to hundreds qubits, which degrades the reliability of NISQ systems.}

Noisy Intermediate-Scale Quantum (NISQ) computers, which have tens to hundreds of quantum bits (qubits) are anticipated to speed up the computational power of classical computers for certain classes of problems. However, the noisy nature of their qubits and operations degrades the reliability of NISQ systems. 
Thus, errors should be mitigated to improve the fidelity of NISQ algorithms, which necessitates error characterization and circuit design approaches to minimize the impact of errors on the quantum circuit output. Randomized benchmarking is typically used to characterize different sources of noise in the quantum hardware including gate, measurement, and decoherence errors~\cite{PhysRevA.85.042311}. The computed error rates are used by noise-aware quantum compilers to generate physical quantum circuits with high fidelity.

While quantum noise is very complex, simplified error models based on error rates collected during the calibration process are typically used for quantum circuit compilation~\cite{{DBLP:journals/corr/abs-1903-10963}}. 
 Depending on the permissible access to the quantum computer, more advanced error characterization can be unavailable. A predictive analysis capable of delivering a projection with a reasonable accuracy about the circuit error rates, despite the lack of complete and accurate error model is also required. A predictive technique, which is computationally fast and applicable to many quantum computing technologies, and takes into account the circuit structure can enhance the compilation process~\cite{10.1145/3408039}.

\begin{figure*}[t]
     \centering
     \begin{subfigure}[b]{1\textwidth}
         \centering
         \includegraphics[width=5.8in]{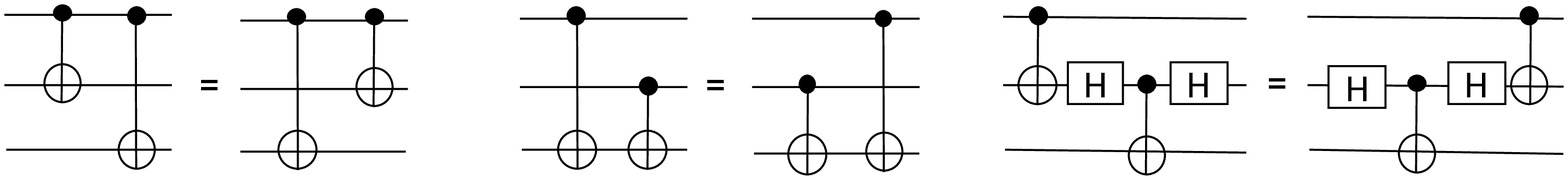}
       \caption{}
         \label{fig:(a)}
     \end{subfigure}
   %  \hfill
     \begin{subfigure}[b]{1\textwidth}
         \centering
         \includegraphics[width=5.8in]{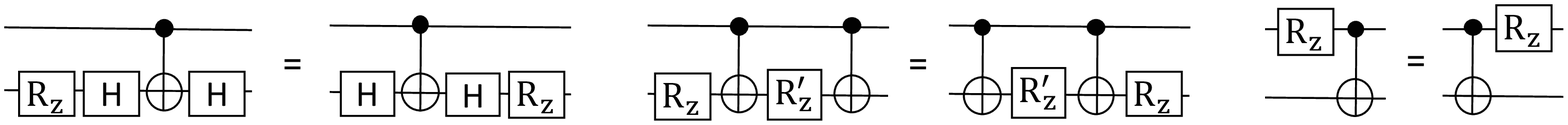}
     \caption{}
         \label{fig:(b)}
     \end{subfigure}
        \caption{Commutation rules of (a) CNOT, and (b) $R_z$ gates~\cite{Optimiz_Q}.}
       % \vspace{-0.05in}
        \label{fig:three graphs}
\end{figure*}

 In this paper, we propose error propagation-based predictive techniques to project the quantum circuit success rate primarily based on gate errors. We take into account the propagation path of different gate errors in the circuit. {\color{black} While the existing estimated success probability typically used by noise-aware quantum compilers to generate physical quantum circuit incorporates the gate error rates into its model, it ignores the order of the gates that don't affect the circuit depth or the gate count.} Our proposed predictive techniques can differentiate between different physical implementations of the same quantum circuit, which use the same physical qubits, share the same set of quantum gates, and have the same depth, but vary in the order of quantum gates in the circuit and their corresponding error rates. {\color{black} We experimentally show the advantage of considering the order of the physical quantum circuit gates in the presence of variable gate errors.} Accordingly, we also propose gate rescheduling algorithms at different quantum circuit abstractions to maximize the fidelity of the quantum circuit. {\color{black} Our proposed approaches can be integrated with other gate rescheduling algorithms that reduce the gate count and the circuit depth.} To the best of our knowledge, no such analysis has been proposed to
 incorporate the impact of the error propagation paths to the quantum gate scheduling procedure. %in the presence of variable gate error rates on the circuit fidelity to the quantum gate scheduling procedure.  %the variation of the gate error rates on the circuit structure
 %optimize the quantum circuit as a major step to maximize the circuit fidelity.asthe following:
 %o the best ofour knowledge, no such analysis has been proposed to predictthe  circuit  error  rates  as  a  major  step  to  optimize  the  circuitfidelity.
 The main contributions of our paper are as follows:
 
 \begin{itemize}
    \item We propose a Weighted Estimated Success Probability (WESP) metric based on the quantum circuit structure. 
    \item We propose new quantum circuit gate rescheduling algorithms at different design levels including complex and elementary gate-level based on our proposed WESP metric to boost the fidelity of the quantum circuit output.
   % \todo{Update: add the QAOA work based on complex gate rescheduling here as well as the abstract.}
    \item We experimentally validate the effectiveness of our proposed metrics and gate rescheduling algorithms using a variety of quantum circuits executed on different IBM quantum computers. %IBM Q16 Melbourne, IBM Q7 Casablanca, and IBM Q5 Santiago quantum computers..
\end{itemize}
 
 %\todo{Update: add a paragraph that talks about the paper organization}

The remainder of paper is organized as follows. Section II provides a background on the quantum circuit compilation, quantum hardware errors, an application of NISQ computers, and different success criteria for quantum circuit evaluation. Section III discusses related works on quantum circuit mapping and gate reordering approaches. Section IV shows the implication of the gate error propagation path on the output state fidelity of the quantum circuit. Section V provides our proposed reliability metric and rescheduling algorithms. Section VI validates the effectiveness of our approaches through several experiments. We conclude the paper in Section VII.

\section{Background}

\subsection{Quantum circuit compilation}
%\todo{where are the references for your discussion. Any thing you talk about should be cited. Professor I cited all the papers}

Quantum algorithms are described using quantum circuits, which are executed on the quantum computer. A quantum circuit comprises of quantum gates, which change the state of the qubits. Single-qubit gates operate on a single qubit such as the Hadamard (H) gate, which creates a superposition state, and the $R_z$ gate, which rotates the qubit around the z-axis. A multi-qubit gate such as Controlled Not (CNOT) entangles two qubits~\cite{nielsen_chuang_2019}. To enable a quantum circuit execution on the targeted NISQ computer, the circuit has to go through several compilation steps. Complex gates should be constructed using elementary gates supported by the NISQ architecture. To reduce the depth and the gate count of the circuit, several gate-level optimization techniques are applied including template matching~\cite{1432873} and gate reordering~\cite{10.1145/3434318} based techniques. In the former one, a cascade of quantum gates is substituted with a sub-circuit with a lower gate count, while in the later one, consecutive gates cancel each other based on the commutation rules of the quantum gates~\cite{Optimiz_Q}, which are defined as follows:
%commutation rules in the quantum computing context is defined as following:
%\todo{please fix the u(x) and V(y) to represent quantum gates that operate in quantum state. if you use the ket notation then you have to explain that it describes the qubit state. Also, update the multiply and use the corresponding operator for quantum gates}
\begin{definition}
Let $U_1$ and $U_2$ be two unitary matrices. $U_1$ and $U_2$ are said to be commutative if $U_1U_2 = U_2U_1$ for any input state. Commutation rules of different quantum gates are shown in Figure~\ref{fig:three graphs}.
\end{definition}
%\todo{I rewrite your paragraph since  there is a lot of similarity with the our previous works (plagiarism). This can cause rejection}
Next, the physical qubits of the quantum circuit are allocated and their gates are scheduled to meet the constraints of the quantum architecture. This process is referred to as quantum circuit mapping. A main challenge in the mapping process is the restricted qubits connectivity in superconducting quantum architectures. Two-qubit gates can be applied to certain pairs of physical qubits described using the coupling graph of the quantum architecture, in which nodes represent qubits and edges show the connectivity between different qubits. To enable arbitrary multi-qubit gates in the presence of the coupling constraint, SWAP operations consisting of three CNOT gates are used for gate scheduling, resulting in an excessive gate count and a large circuit depth. Efficient mapping approaches are used to generate physical quantum circuits with as a minimum gate count as possible to be executed on the quantum computer. These approaches often represent the quantum circuit {\color{black}as a directed acyclic graph, referred to as a} gate dependency graph, where each node represents a gate and each edge represents a direct dependency between two gates. {\color{black} The dependency graph is divided into levels, where each level consists of quantum gates applied simultaneously in the same circuit layer. All the nodes reachable from a given node $i$ in the graph are considered as the gates which are dependent on the $g_i$ gate ($g_i$ reachable gates).}

\begin{figure}[t]
\vspace{-0.05in}
     \centering
         \includegraphics[width=3.3in]{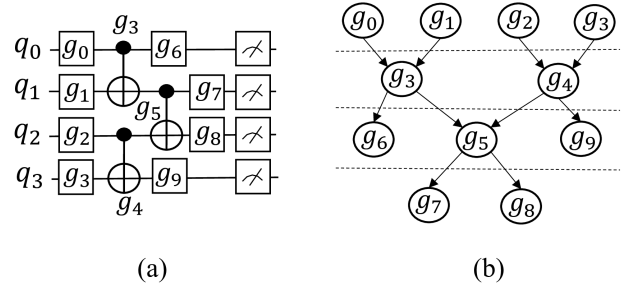}
         \caption{{\color{black}(a) A quantum circuit and (b) its corresponding gate dependency graph.}}
         \label{fig:gate_dep_general}
     
\end{figure}

{\color{black}
\begin{example}
 Figure~\ref{fig:gate_dep_general} shows an example of a quantum circuit and its corresponding gate dependency graph in which the gates in each circuit layer are mapped to the graph nodes applied at the same level. For each quantum gate, the list reachable gates is extracted from the gate dependency graph. For example, there are 3 reachable gates from $g_3$ gate, which are  $g_5$, $g_6$, $g_7$, and $g_8$ gates.
\end{example}
}

\subsection{Reliability of quantum devices}

\begin{figure}[b]
     \centering
    \vspace{-0.1in}
     \begin{subfigure}{.35\columnwidth}
         \centering
         \includegraphics[width=0.8in]{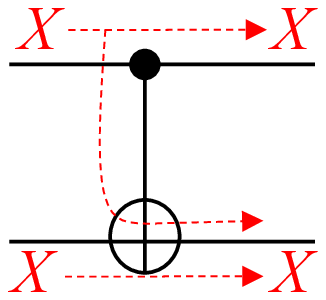}
         \caption{}
         \label{fig:(a)}
     \end{subfigure}
   %  \hfill
     \begin{subfigure}{.35\columnwidth}
         \centering
         \includegraphics[width=0.8in]{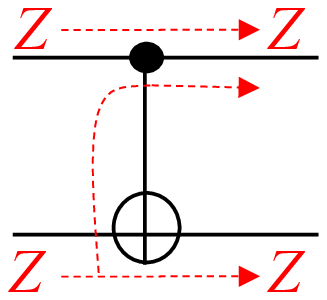}
         \caption{}
         \label{fig:(b)}
     \end{subfigure}
        \vspace{-0.05in}
        \caption{The propagation path of (a) X and (b) Z Pauli gate errors through a CNOT gate.}
     %   \vspace{-0.05in}
        \label{fig:pauli}
\end{figure}

Different sources of noise contribute to the noise complexity of NISQ systems. The elevated noise levels endanger the reliability of quantum circuits. Gate errors are a major source of quantum computing noise. They are modeled as Pauli gates that operate on a single qubit. They are represented as $X$ (a $\pi$ rotation around the x-axis), $Y$ (a $\pi$ rotation around the y-axis), $Z$ (a $\pi$ rotation around the z-axis), and $I$ (identity matrix) single-qubit gates~\cite{janardan2016analytical}. The propagation of some of the Pauli errors through the CNOT gate is shown in Figure~\ref{fig:pauli}. The measurement/read-out error can occur while measuring the output state of the quantum circuit. A qubit can lose its state after a period of time, resulting in decoherence errors. Quantum gates that act on different qubits simultaneously can cause crosstalk errors. 
%\todo{send me the ppt of this figure. The resolution is bad}
%\todo{we need to show how pauli error propagate through clifford gates. It is explained in the reference above. To show this we need to draw the figure that shows error propagation. Please draw it in your own way son it doesn't look like we are copying it from the paper}

As NISQ computers do not support quantum error correction, they often rely on noise-aware quantum compilers to mitigate errors. The quantum compiler should maximize the success probability of the physical quantum circuit. An Estimated Success Probability (ESP) can be used by a noise-aware quantum compiler to
select the physical qubits of the quantum circuit and schedule the quantum gates. Given the  gate ($e_{g_i}$) and measurement ($e_{m_i}$) errors for all the quantum circuit gates, and its qubits, ESP is computed as $\textstyle \prod^{G-1}_{i=0}(1 - e_{g_i}) \times \prod^{Q-1}_{i=0}(1- e_{m_i})$~\cite{DBLP:journals/corr/abs-1903-10963}. {\color{black} Randomized Benchmarking sequences are executed on the quantum hardware to compute single- and two-qubit gate errors based on the average sequence errors. The qubit measurement/readout error is computed as the average measurement error 0 and 1, in which the qubit is initially assigned to state 1 and 0, respectively. 
}
%Modeling of error is not supported by NISQ devices. Hence the quantum circuit has to be compiled in the presence of noise, and the output has to be measured to analyze the amount of noise in the circuit. We adopt Estimated Success Probability (ESP), defined as the summation of the product of gate errors of the quantum circuit based on the physical constraint of the device. There are several noise-aware compilers that try to minimize the impact of the error in the quantum device.

%\todo{Since a new paragraph that focus on quantum circuit compilation in the presence of noise you need one opening sentence that explain that. Remember the first sentence should give an idea about what we need to talk about in the remaining paragraph} 

%\todo{take PSt definition from here and add it at the end of section II.A when you finish the discussion about mapping. You can emphasize that the once you generate the physical quantum circuit it is executed on NISQ computer. To assess the output reliability of the circuit, different success criteria can be applied. We adopt Probability of Successful Trails (PST) ..... (define it here) }
%different success  and the results are obtained 
\subsection{Quantum approximation optimization algorithm}
Quantum approximation optimization algorithm (QAOA) is a hybrid quantum-classical algorithm for solving optimization problems. It relies on not only the output of the quantum circuit but also the classical optimizer that updates the circuit parameters to improve the solution of the optimization problem~\cite{farhi2014quantum}. The objective of the algorithm is to minimize/maximize a cost function (C(X)) of a given problem described using the following Hamiltonian: $H= \sum_{x\in{0,1}^n}{C(x)\ket{x} \bra{x}}$. A QAOA quantum circuit contains two main components, namely, phase separation and mixing operations, which are applied $p$ times repeatedly with ($\gamma_1$,$\beta_1$), ..., ($\gamma_p$,$\beta_p$) parameters. %,$\gamma_p$, $\beta_1$, ..., $\beta_p$ parameters. 
The phase separation can be represented using ZZ gates, while mixing operations are represented using single-qubit gate rotation around the x-axis (R$_x$). An example of a problem to be solved using the QAOA algorithm is the Maximum Cut (Max-Cut) problem. For a graph with n vertices, the solution of the Max-Cut problem divides the graph into two subsets such that the total weight of the edges between those subsets is as maximum as possible. The graph vertices are described as a string of n qubits in which each qubit can be measured as 0 or 1. The QAOA circuit output provides a candidate solution of the Max-Cut problem. Using an iterative procedure, we can find the n-qubit string that maximizes the total weight of the edges between the two subsets, referred to as the cost function. 

\subsection{Quantum circuit success metrics}
Different success criteria/metrics have been proposed to show the impact of the quantum compilation approaches on the quantum circuit output. For a single-output quantum circuit, the Probability of Successful Trails (PST) measures the probability of the correct output of the quantum circuit as $\frac{\text{Number\:of\:Successful\:Trials}}{\text{Total\:Number\:of\:Trials}}$~\cite{Tannu:2019:QCE:3297858.3304007}.
For hybrid quantum algorithms, application-specific success criteria are used. For example for QAOA, which maximizes or minimizes a cost function, the Approximation Ratio (AR) of the cost function quantifies the success of the quantum circuit~\cite{9251960, PhysRevX.10.021067}. It is defined as $\frac{\text{Mean\:of\:Cost\:Function\:Over\:all\:Sampled\:Output}}{\text{Maximum\:Cost\:Function\:Value}}$. % the mean of the cost functoin of all the sampled ouputs of the quantum circuit divided by
 Given AR, we can compute the Approximation Ratio Gap (ARG) to quantify how close is the output of the actual execution of QAOA circuit to the simulation result~\cite{9251960}. It is computed as $\frac{\text{AR\:of\:Simulation\:-\:AR\:of\:Execution}}{\text{AR\:of\:Simulation}}\times100$. The smaller the gap, the more reliable the QAOA circuit.

%\balance
%\newpage

\section{Related work}
\label{related_work}
%\todo{swap gate should be replaced with swap operation everywhere}
\subsection{Quantum circuit mapping}
%Mapping logical quantum circuits to the quantum architecture based on the physical constraints 
Since quantum circuit mapping is an NP-Complete problem~\cite{siraichi2018qubit}, traversing the search space is speculative for quantum architectures with a large number of qubits. Various heuristics-based techniques have been proposed to address this problem~\cite{9384317}. Some of these techniques target minimizing the number of circuit SWAP operations~\cite{itoko2019optimization,8702439,wille2019mapping,zulehner2018efficient,10.1145/3297858.3304023}, while others focus on minimizing the circuit error rates~\cite{Tannu:2019:QCE:3297858.3304007,Ash-Saki:2019:QQR:3316781.3317888}. % of the quantum architectures to improve the success probability of the quantum circuit output. 
For example, quantum circuit mapping can be achieved by using BRIDGE and SWAP operations as transformation rules inserted based on a dynamic programming or a heuristic-based look-ahead schemes~\cite{itoko2019optimization}. The BRIDGE gates are composed of a sequence of four CNOT gates. %that replace the CNOT gate between two non-adjacent qubits.
%\todo{still the main idea of this paper is missing which is the how?}%\todo{if we add 4 cnot will obviously execute 4 cnot. do we need to add results xxxx} %, whereas SWAP gates only swap two qubits. 
SAT solver minimizes the number of SWAP operations in the physical quantum circuit at the cost of extensive computation overhead (e.g.~\cite{wille2019mapping,8702439}). The proposed technique in~\cite{8702439} divides the quantum circuit into sub-circuits that require no SWAP operations within each sub-circuit. %\todo{what is the main property of these sub-circuit? require no swap within each sub-circuit?}
Next, the SAT solver is used to construct the quantum circuit by inserting a minimum number of SWAP operations between the sub-circuits that share qubits. %\todo{where do you apply SAT? in step 1 or 2?} \todo{rewrite}
In~\cite{wille2019mapping}, a logical quantum circuit is mapped to a physical quantum circuit using a SAT solver too. % by replacing the CNOT gates with SWAP operations for non-adjacent qubits and inserting H operations where the control and target qubits needed to be switched. 
Constraints are added to the SAT solver based on the coupling graph of the quantum architecture to reduce its complexity at the expense of the gate count. %~\todo{Still this problem holds: Again, the sentence that you wrote doesn't differentiate this paper from the one that divides the circuit into sub-circuit. both use sat solver. So, please write something that differentiates this approach from the previous one}
%An A* algorithm was proposed to map the quantum circuits to the IBM QX architecture~\cite{zulehner2018efficient}.%\todo{Not correct. this paper didn't consider the error rates. This can be an easy reason for the paper to get rejected. Please don't write any sentences before you check the paper. They were considering only the number of swaps}.
%\todo{move the reference always to the end of the sentence unless you are talking about more than one reference in a sentence}  
%In this approach, the quantum circuit is divided into layers of CNOT that can be applied in parallel. 
An A* algorithm explores the search space in the coupling graph to find the shortest path between qubits for each two-qubit gate, and thus, reduce the number of SWAP operations in the quantum circuit~\cite{zulehner2018efficient}. %\todo{re-write: A* explore the search space to find the shortest path between each qubit pair. check the paper and write a reasonable explanation according to it}
%A SWAP-based heuristic approach is proposed to optimizes the qubit allocation by comparing the initial mapped quantum circuit with the mapping of the reversed quantum circuit~\cite{10.1145/3297858.3304023}. 
To address the difference in the error rates of various qubits, a mapping technique has been proposed based on Dijkstra algorithm, which schedules quantum gates and allocates physical qubits with low error rates~\cite{Tannu:2019:QCE:3297858.3304007}. %\todo{NAMES don't matter. What do they do? use greedy approach? DFS or BFS? check}
%Mapping of physical qubits is improved to enhance the output success probability of the quantum circuit \todo{why, for example, here? both this paper and the previous one use error rate, so why do you add, for example, to the second sentence?}, 
The success probability of the physical quantum circuit can be further improved by searching for isomorphic sub-graphs in the quantum architecture's coupling constraint with the highest ESP~\cite{Ash-Saki:2019:QQR:3316781.3317888}. Other mapping approaches address decoherence, correlated, and unexpected errors (e.g~\cite{8824907,Tannu:2019:EDM:3352460.3358257,10.1145/3400302.3415684}). %, which cause temporal variations in the qubit error rates by minimizing the circuit depth (e.g~\cite{8824907}) %\todo{add a reference} or combining the output distribution of various qubit allocations~\cite{Tannu:2019:EDM:3352460.3358257}. %(e.g.,~\cite{Tannu:2019:EDM:3352460.3358257,10.1145/3373376.3378477}). Errors might occur when the same quantum circuit is executed multiple times. To minimize the impact of correlated errors, the authors proposed a wide range of mapping the quantum circuits by allocating different physical qubits for different runs~\cite{Tannu:2019:EDM:3352460.3358257}. The incorrect output can be suppressed by combining the output success probability of various qubit allocations.\todo{did you check this reference? you shouldn't be adding it here. You should be talking about diverse ensabmble. You really need to go quickly over each paper that you cite (open in IEEEXplore)}
To mitigate crosstalk errors, they should be characterized first, followed by an efficient mapping approach that selectively serializes circuit gates, while satisfying the required coherence time~\cite{10.1145/3373376.3378477}. 
%the authors propose different methods to characterize the impact of crosstalk errors and exploit the error rates in developing an efficient mapping approach based on a metric that measures the total crosstalk effect as the sum of occurrences of CNOT pairs that are adjacent in each layer~\cite{10.1145/3373376.3378477}.\todo{rewrite: to mitigate crosstalk errors .... the authors propose different methods to characterize the impact of crosstalk errors and exploit the error rates in developing efficient mapping approach based on xxxxx } %The crosstalk of the qubits on the quantum architecture is high for adjacent qubits. The characterization of crosstalk can be made more effective using this metric.

%\hl{stopped here***}

\subsection{Gate reordering of quantum circuits}
A quantum circuit consists of layers of gates that can be applied in parallel. Consecutive commuting gates can be reordered while preserving the state of the quantum system~\cite{Optimiz_Q, Guerreschi_2018}. The commutation rules of the gates have been utilized to reduce the quantum circuit depth, gate count~\cite{9218558,10.1145/3400302.3415620,Optimiz_Q, Guerreschi_2018}, and optimize the control pulses~\cite{Shi_2019}. A heuristics-based approach has been developed to perform a gate-level optimization by reordering quantum gates to reduce the circuit gate count~\cite{Optimiz_Q}. %In this approach, the gates are reordered without affecting the logic of the quantum circuit\todo{if you are reordering Clifford gates how do you reduce rz gate count? rz is not a Clifford gate}.
A two-step approach has been proposed to reduce the depth of the quantum circuit by constructing a dependency graph based on the circuit gates and then swapping the gates that satisfy the commutation rules to reduce the circuit depth~\cite{Guerreschi_2018}. %\todo{is the approach reducing run time of the mapping or reducing the depth of the quantum circuit. CHECK} \todo{Are you sure that we construct depedncy graph of the first layer only????} \todo{what do you mean by rotation operations???? Again not clear where the reordering or commutative operations are used}\todo{a graph without coupling constraint seems weird. Are you talking about the gate dependency graph? also i still don't understand the algorithm here. Most of the approaches above use dependency graph } 
A compilation approach has been developed to enhance the performance of QAOA circuits by reordering the complex gates~\cite{9218558,9251960}. Due to the commutation relation, most of the dependencies are ignored. %Instead, a two-level search process is adopted by considering minimal depth of the circuit as an objective and use a breadth-first search algorithm to find the possible two-layer interchanges in each iteration until the desired depth is achieved. 
Instead, a two-level search process is adopted to minimize the circuit depth using a breadth-first search algorithm, which finds a possible two-layer interchanges in each iteration until the desired depth is achieved. 
%\todo{vedika when you edit something read the entire sentence and see if it makes sense to you. The previous sentence is broken}
%\todo{this is ICCAD paper. Find the citation from ICCAD and update it. Don't use arxiv citation}, 
A layout synthesis algorithm has been proposed to minimize the circuit depth by reordering  %\todo{complex?}
complex gates of QAOA circuits using a Satisfiability Modulo Theories (SMT) solver~\cite{10.1145/3400302.3415620}. 
%\todo{do they consider commutative rule or only reorder complex gates in qaoa. please check and fix} % is used to minimize the depth of the QAOA circuit while scheduling the gates.
An optimized pulse level compilation approach has been proposed to address the inefficiency of the standard gate compilation process that directly translates the logical instructions to control pulses~\cite{Shi_2019}. In this approach, the gates are reordered based on commutation rules, and then a small set of gates are aggregated to a larger operation by finding the optimal control pulse based on the gradient descent method. %\todo{check: is it the delay in the compilation or the circuit. verify from the paper. also I think the goal is related to the pluse not to reduce any delay. CHECK THE PAPER} 
%\todo{You need to fix the next paragraph. First, don't start with Other gate reordering approaches. it is one approach. Use the same flow as the previous sentences. I rewrite it. Please check this and try to learn:Gate reordering has also be applied to minimize the impact of coherence errors~\cite{9296804}. Faults in the form of additional gates due to coherence errors are injected into the quantum circuit to analyze their impact on the circuit output. Accordingly gates are reorder at circuit locations with coherence errors that significantly affect the circuit output.}
 Gate reordering has also been applied to minimize the impact of decoherence errors~\cite{9296804}. Faults in the form of additional gates due to decoherence errors are injected into the quantum circuit to analyze their impact on the circuit output. Accordingly, gates are reordered at the circuit locations, in which the decoherence errors significantly affect the circuit output.
 {\color{black}
 In this work, we propose additional compilation layers applied to the physical quantum circuit post-mapping. We restrict the search space of our gate rescheduling algorithms to maintain the gate count and the circuit depth of the optimized quantum circuit using the previously proposed compilation approaches. Thus, our work can be incorporated with the other compilation/optimization layers to further improve the output state fidelity of the quantum circuit.  
 }
%Other gate reordering approaches are recently proposed based on noise insertion on qubits to minimize the coherence error~\cite{9296804}. The noise is inserted into a single qubit for a single execution, and the impact is evaluated at the output of the quantum circuit. For n qubits in a quantum circuit, the noise is injected and executed for n times. The impact is analyzed by plotting the output probability of the quantum circuit using the heatmap.
%\bibliographystyle{unsrt}

\begin{figure}[t]
     \centering
     \includegraphics[width=3.4in]{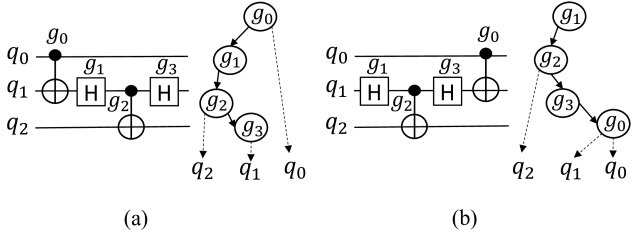}
     \caption{(a) A quantum sub-circuit and its gate dependency graph, and (b) its corresponding equivalent sub-circuit and its gate dependency graph.}
       % \vspace{-0.05in}
        \label{fig:pro_example}
\end{figure}

\begin{figure*}
     \centering
     \includegraphics[width=6.9in]{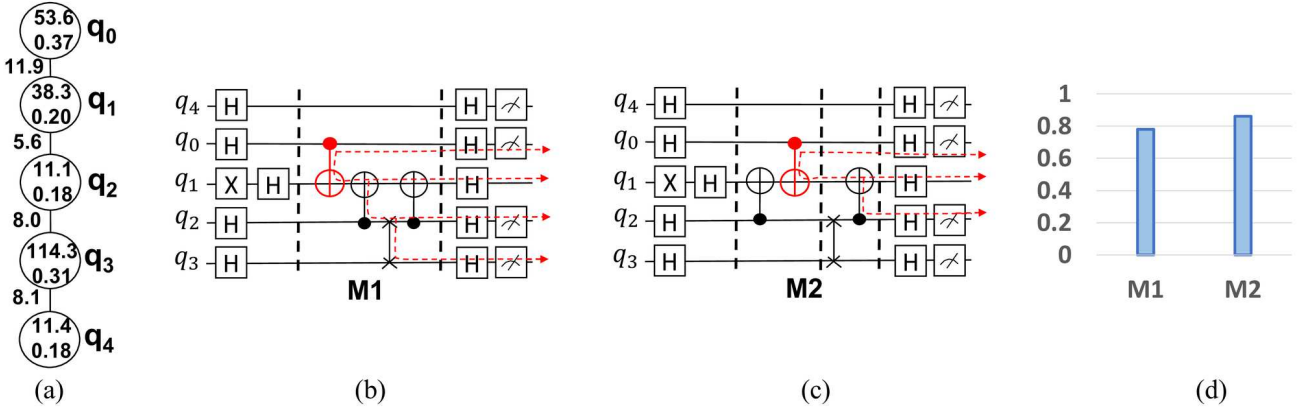}
     \caption{(a) IBM Q Santiago architecture, 4-qubit BV implemented using (a) M1 and (b) M2 mappings,  and (d) their corresponding PST value.}
       % \vspace{-0.05in}
        \label{fig:mapping}
\end{figure*}

%\todo{fix figure (d). There is additional space around the figure. Also convert percentage to probability}
%\section{Implication of the quantum circuit structure on the output state fidelity}
\section{Implication of the gate error propagation path on the output state fidelity}
%\todo{move hadamard gate before the barrier in both (a) and (b) of fig 2}

%\todo{what is ESP of this circuit? please compute it and add it. The example should emphasize that different gate reordering gives different outputs but still ESP is the same.}
Stochastic Pauli noise is a primary factor to estimate the circuit error rates~\cite{{DBLP:journals/corr/abs-1903-10963}}. While tracking the precise impact of the error propagation on the circuit can be computationally expensive %is hard
in the presence of universal %(non-Clifford) 
gates~\cite{janardan2016analytical}, 
two-qubit gates enable error propagation across different qubits. Thus, the location of the quantum gates especially the two-qubit gates can affect the circuit error rates. {\color{black} For the sake of simplicity}, we assume  that the two-quit gate (CNOT) error impacts the control and the target qubits, and thus, propagates to the gates applied next to the control and the target qubits. 

%\todo{remove figure 4 and 5. I added figure 5 to show how to add the dotted line to show the error propogation path. Please add the same path to fig 2(a) and fig 2(b). Also check the quality of the figure. You should save it as jpg first and not png and then convert to pdf then convert to eps. Please improve the qualtiy of figure 2}

\begin{example}
To illustrate the gate error propagation through the quantum circuit, {\color{black}let's consider} the example provided in Figure~\ref{fig:pro_example}. Two equivalent sub-circuits using the commutation rules of quantum gates and their corresponding gate dependency graphs are provided. {\color{black} We use a dotted line to show the output qubits affected by the gates in the dependency graph.}
Every gate in the sub-circuit can produce Pauli errors that propagate to the qubit output through the gate dependency graph~\cite{janardan2016analytical}. In Figure~\ref{fig:pro_example}(a),
$q_0$, $q_1$, and $q_2$ output qubits are affected by $\{g_0\}$, $\{g_0, g_1, g_2, g_3\}$, and $\{g_0, g_1, g_2\}$ gate errors, respectively.
On the other hand, in Figure~\ref{fig:pro_example}(b), $q_0$, $q_1$, and $q_2$ output qubits are affected by $\{g_0, g_1, g_2, g_3\}$, $\{g_0, g_1, g_2, g_3\}$, and $\{g_1, g_2\}$ gate errors, respectively. Thus, while the two quantum sub-circuits in Figure~\ref{fig:pro_example} are equivalent, their output qubits $q_0$, and $q_2$ are susceptible to different sets of gate errors.
%errors caused by $\{g_0\}$, $\{g_0, g_1, g_2, g_3\}$, and $\{g_0, g_1, g_2\}$ are propagated to $q_0$, $q_1$, and $q_2$ output qubits, respectively. On the other hand, in Figure~\ref{fig:pro_example}(b), errors caused by $\{g_0, g_1, g_2, g_3\}$, $\{g_0, g_1, g_2, g_3\}$, and $\{g_1, g_2\}$ are propagated to $q_0$, $q_1$, and $q_2$ output qubits, respectively. Thus, while the two quantum sub-circuits in Figure~\ref{fig:pro_example} are equivalent, their output qubits $q_0$, and $q_2$ are affected by different set of gate errors.
\end{example}

Due to the spatial variation in error rates of the quantum hardware, gates applied to different qubits result in different error rates. Although all gate errors affect the fidelity of the quantum circuit, the two-qubit gate errors has lower fidelity than single-qubit gate errors. %Furthermore, two-qubit gate errors are typically higher than single-qubit gate errors. %Furthermore, two-qubit gate errors them self applied to different pair of qubits can have distinct error behavior. 
Thus, reordering quantum gates while maintaining the circuit functionality changes not only the error propagation paths but also the gate error impact on the output state of the circuit. In the previous example, if $g_0$ gate has the highest error rate in the circuit, which significantly deviates from other gate errors in the circuit, applying $g_0$ gate at the first layer of the sub-circuit as in Figure~\ref{fig:pro_example}(a) will spread $g_0$ error to all the output qubits unlike the equivalent sub-circuit in Figure~\ref{fig:pro_example}(b) in which $g_0$ error affects only $q_0$ and $q_1$ output qubits.

%\todo{in fig 2. add the coupling constraint first followed by the two BV circuits}
%\todo{are the errors in the coupling graph multiplied by 10E-3 or 10E-2? Professor, the values are all multiplied by 10E-3. YOU should modify the figure to show}

\begin{example}
Figure~\ref{fig:mapping} shows the impact of gate reordering on the quantum circuit output fidelity. A 4-qubit Bernstein–Vazirani (BV) circuit is mapped to satisfy the coupling constraint of IBM Q Santiago architecture.  Figure~\ref{fig:mapping}(a) represents the coupling graph and the error rates of IBM Q Santiago architecture, in which the values in the upper and lower part of each node represent the read-out and single-qubit errors, respectively, while the edge label is the two-qubit gate error applied to the corresponding pair of qubits. All error rates are multiplied by $10^{-3}$. Two physical implementations of the BV quantum circuit, namely M1 and M2, are provided in Figure~\ref{fig:mapping}(b) and~\ref{fig:mapping}(c), respectively. The difference between the two physical implementations of the circuit is the location of the first two CNOT gates of the quantum circuit, which can be reordered based on  the commutation rules. The CNOT gate highlighted in red color has the highest two-qubit gate error rate. According to the order of the quantum gates, the error propagation path also changes as shown in Figure~\ref{fig:mapping}(b) and~\ref{fig:mapping}(c). We demonstrate the impact of the gate reordering on the quantum circuit output by executing both M1 and M2 quantum circuits on IBM Q Santiago architecture, and reporting their PST as shown in Figure~\ref{fig:mapping}(d). To eliminate the impact of gate reordering on decoherence errors, we insert barriers before and after applying all the two-qubit gates in the two circuit implementations in addition to a barrier after the second CNOT gate of M2 quantum circuit in Figure~\ref{fig:mapping}(c), which ensure that the qubits in the two implementations are used for the same period of time. The PST values are $0.784$ and $0.861$ for the BV quantum circuit generated using the M1 and M2 mapping, respectively, and executed for 8192 shots/trials. Our observation indicates that reordering gates with higher error rates to earlier circuit layers will result in error propagation to larger number of circuit gates, and thus, degrade the output state fidelity. While PST is different for both mapping, the ESP of both M1 and M2 mapping is $0.7757$, which implies the need to consider the gate error propagation path for estimating the success rate of the circuit, and thus, refining the quantum circuit mapping policies.

\end{example}

\iffalse
\todo{
First you need to emphasize that with a given quantum circuit mapping and the same depth there is a room to reoder gates while maintaining the circuit depth. You take a simple BV example with 4 qubits and you show that you can reorder the CNOT gates in different ways and show the corresponding PST for each reordering. Make sure you add a barrier after the single qubit gate at layer one and before the single qubit gates at the last layer to insure that the circuit depth is the same. This motivates the need of taking the circuit structure into account.

Then, you need to explain that the location of the gate in the circuit impacts the error propagation path of that gate errors. THus, applying noisy qubit gates at the beginning of the circuit will increase the impact of the errors in the circuit. Here you also take the same BV example and show for a very noisy CNOT gate applies at the beginning its error propagation path as arrows through the circuit (in red color). You can give also red color to the CNOT gate with the highest error (make sure that there is a significant difference in the different CNOT error gates). Use the same BV example as before. You conclude this section with the need to take location of very noisy gates in the circuit as they may spread errors aggressively thought the circuit.

Next section you can start talking about your reliability model WESP using the same example as before to demonstrate how to compute it.
}
\fi
\section{Error propagation-based gate rescheduling}

\subsection{Weighted estimated success probability}
\begin{figure}
     \centering
  \begin{subfigure}[b]{0.49\textwidth}
         %\centering
         \includegraphics[width=3.4 in]{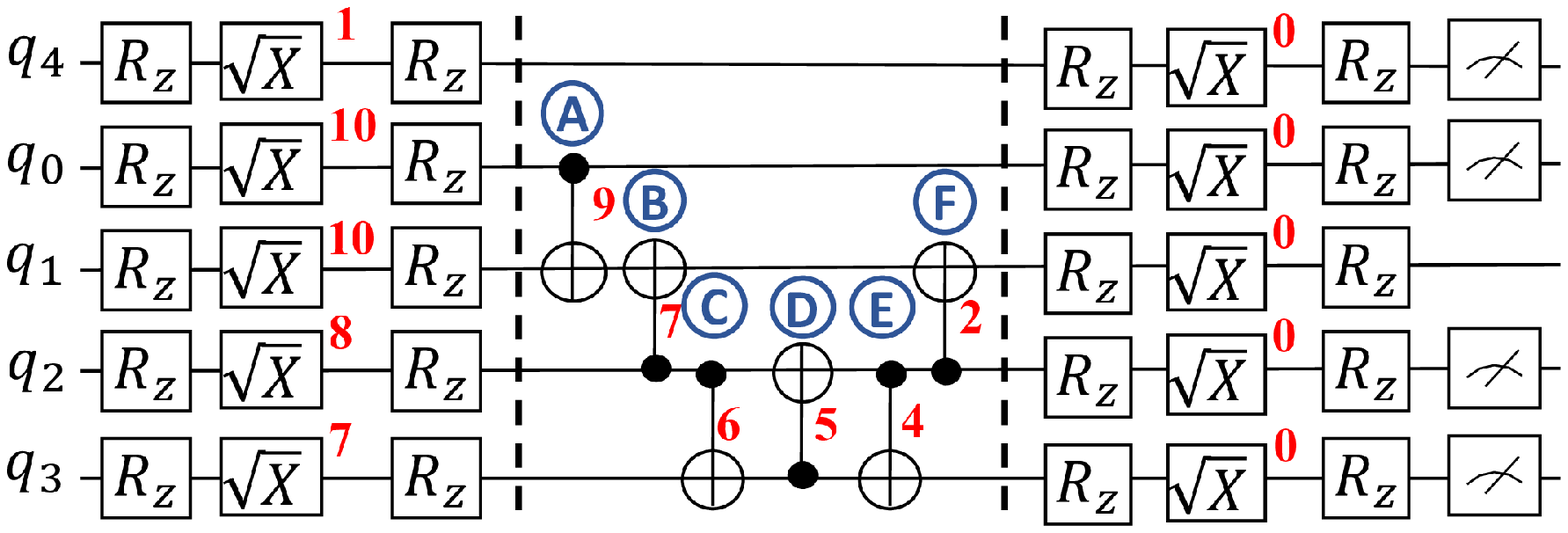}
         \caption{}
         \label{fig:m1wesp}
     \end{subfigure} \vspace{1mm}
     %\hfill
     \begin{subfigure}[b]{0.48\textwidth}
         \centering
         \includegraphics[width=3.4 in]{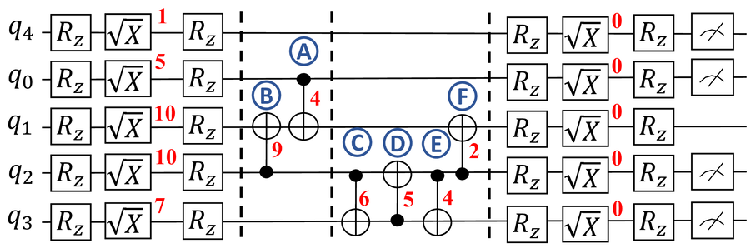}
         \caption{}
         \label{fig:m2wesp}
     \end{subfigure}%\hspace{1mm}
        \caption{{(a) The number of reachable gates ($S_{i}$}) for each quantum gate in M1, and (b) M2 mappings.}
       %\vspace{-0.05in}
        \label{fig:wesp}
\end{figure}
%\todo{talk about the gate dependency graph that is used to compute the number of dependent gates }
ESP enables quick and easy estimation of the circuit success probability. Yet, it overlooks the circuit structure since different orders of the quantum gates yield the same ESP. 
To account for the penalty of rescheduling the same set of physical quantum gates with variable error rates on the output state fidelity of the quantum circuit, we modify the ESP metric and propose a new metric referred to as Weighted Estimated Success Probability (WESP) defined as 
%To analyze the performance of the quantum circuit after reordering the gates based on commutative property we propose a new metric weighted estimated success probability (WESP) defined as 
\begin{equation*}
\textstyle
WESP =\prod^{G-1}_{i=0}(1 - (\lambda_{i} + e_{g_i})) \times \prod^{Q-1}_{i=0}(1- e_{m_i})\:\:\end{equation*}
$\lambda_{i}$ tunes the error rate of the corresponding $g_i$ gate to account for the gate position as
\begin{equation*}
\lambda_{i} = w_i(e_{g_i}- {\color{black} \min(E_{G})}),
\end{equation*} where $w_i$ and {\color{black} $\min(E_{G})$} are the weight of the $i^{th}$ gate and the minimum gate error rate in set of gate errors {\color{black}($E_{G}$)} of the quantum circuit, respectively. Instead of multiplying the weight of each gate by its error rate, which can significantly reduce the gate error of very noisy gates applied in later layers of the circuit, we multiply the weight with the difference between the current gate error and the minimum gate error in the circuit {\color{black} to account for the variation in the gate errors}. The resulting number is added to the gate error. To compute $w_i$, the gate dependency graph of the circuit is constructed first. For each gate ($g_i$) (node) in the circuit (graph), we identify the number of all subsequent dependent gates (reachable gates) that are applied to the same qubit or other qubits that are directly or indirectly connected to the current gate through two-qubit gates ($S_i$). For a quantum circuit with $G$ gates, the weight of the $i^{th}$ gate ($w_i$) is computed as $w_{i}=\frac{S_i}{G}$.
%${w_i} = \textstyle \frac{w_{g_i}}{N}$, ${w_{g_i}}$ represents fan-out of each gate in the quantum circuit and ${N}$ is the total number of gates in the circuit, $G$ and $Q$ are the number of gates and qubits with measurement operation, while $e_{g_i}$, $e_{m_i}$ and $min_{e_{g_i}}$ are gate, measurement and minimum of gate error rates of the quantum circuit, respectively. Fan-out is defined as the number of input gates driven by the output of previous gates. 
We exclude the $R_z$ gate since the error rate is zero. Thus, the term $\lambda$ is used to approximate the impact of noisy quantum gates on the output state fidelity of the quantum circuit given the number of subsequent dependent (reachable) gates. To simplify our analysis, we omit how Pauli errors propagate through different quantum gates in the circuit. Instead, we assume that for each gate, Pauli errors will propagate to its corresponding qubits.

%\todo{New: please fix this figure. It should in the center. There is additional space around the figure}

\begin{example}
Figure~\ref{fig:wesp}(a) and (b) show the physical quantum circuits implemented using elementary gates supported by IBM quantum computers for M1 and M2 quantum circuits in Figure~\ref{fig:mapping}, respectively. The $S_i$ values for each single- and two-qubit gates are shown in red color. For example, in the physical circuit provided in Figure~\ref{fig:wesp}(a), the $\sqrt X$ gate applied to qubit $q_4$ in layer 2 is connected to the next $\sqrt X$ gate applied to the same qubit at a later layer in the gate dependency graph. Hence, the value of $S_i$ for the first $\sqrt X$ gate is 1. Similarly, the value of $S_i$ for all the other gates except $R_z$ gates are computed. The number of erroneous gates ($G$) is 16. Based on the device error rates as shown in Figure~\ref{fig:mapping}(a), the value of WESP for M1 mapping is $0.7748$. On the other hand, WESP for M2 mapping is $0.7764$, which indicates that M2 circuit provides a higher PST. This observation is aligned with the PST value of M1 and M2 circuits provided in Figure~\ref{fig:mapping}(d). %and ESP for both mapping is $0.825585$. The PST for m1 and m2 is $54\%$ and $71\%$.

%\hl{stopped here}
\iffalse
Lets consider the quantum circuits with M1 mapping as shown in Figure~\ref{fig:m1wesp}. The ${w_{g_i}}$ values for each single and two-qubit gates are highlighted in red, and the value of $N$ is 16. For example, the output of  $\sqrt X$ gate in qubit $q_0$ is feed as input to the next $\sqrt X$ gate in the same qubit, hence the value of ${w_{g_i}}$ for the first gate is 1. Similarly, the value of  ${w_{g_i}}$ for all the gates except $R_z$ gate is computed. Based on the device error is shown in Figure~\ref{fig:mapping}(c), the value of WESP for $M_1$ mapping is 0.817531.
\fi
\end{example}

\subsection{Elementary gate rescheduling based on WESP}
We exploit the gate commutation rules to reduce the impact of Pauli errors while maintaining the depth of the circuit. We propose a gate 
 rescheduling algorithm post-quantum circuit mapping guided by the proposed WESP metric as a new optimization layer to reduce the circuit gate errors. Since the gates of the physical quantum circuit satisfy the coupling constraint of the quantum hardware, no additional SWAP operation is required. We develop a greedy approach for gate rescheduling based on the circuit dependency graph ($D(V,E)$) as described in Algorithm~\ref{alg:alg2}. The objective of our rescheduling algorithm is to minimize WESP while maintaining the circuit depth ($Depth_C$) to avoid any additional decoherence errors. An important condition is added to ensure that the updated gate scheduling does not increase the depth of the circuit ($No\_Inc\_Depth$). For $g_i$ gate operating on $q_x$ and $q_y$ qubits at layer $t$ and $g_j$ gate operating on $q_x$ and $q_z$ qubits at layer $t+k$, $g_i$ and $g_j$ can be reordered if they are commuting and $q_y$ and $q_z$ are {\color{black}idle} at layers $t+k$ and $t$, respectively. Our evaluation is repeated for every pair of {\color{black} immediate} dependent gates in the gate dependency graph. For a quantum circuit with $D(V,E)$ gate dependency graph, in which $V$ (nodes) represents the {\color{black}set} of quantum circuit gates and $|E|$ (edges) shows the dependency between different gates, the time complexity of calculating WESP is {\color{black}$O(|V|+|E|)$}. Thus, the time complexity of our greedy heuristic, which is provided in Algorithm~\ref{alg:alg2} is {\color{black}$O(|V|^{2} \cdot (|V|+|E|))$}. Our proposed gate rescheduling approach is computationally fast compared to the exhaustive approach that requires checking the entire search space, which can be very large for quantum circuits with large number of commuting gates.

\begin{example}
To motivate the need and the effectiveness of our greedy approach, we exhaustively generate all possible ways of gate scheduling for our BV circuit provided in Figure~\ref{fig:mapping} and~\ref{fig:wesp}. We pick the circuit with the highest WESP and compare it with the generated circuit using our greedy approach in terms of WESP and PST. Specifically, we label each CNOT gate in Figure~\ref{fig:wesp} to list all possible combinations of gate scheduling. The total number of all possible ways of gate scheduling, which maintain the depth of the circuit is 14. Among all the gate scheduling combinations, B-C-D-E-F-A has the highest WESP of 0.7768, while the resulting circuit based on our greedy algorithm has a CNOT combination of B-A-C-D-E-F with a WESP of 0.7764. The PST value of the quantum circuit based on the exhaustive and our greedy approach are 0.863 and 0.861, respectively. Accordingly, the gap between the exhaustive and the greedy solutions is very small.
\end{example}

\begin{algorithm}%[H] This H remove space before and after the algo
\SetAlgoLined
%\SetLine
\begin{small}
{\color{black}
\caption{Gate rescheduling} 
\label{alg:alg2}
% \begin{algorithmic}[1]
\KwIn{$C$ = The physical quantum circuit}
\KwOut{$C^{'}$ = The updated physical quantum circuit}
Construct the $D$ graph of $C$; \\Initialize $C^{'}$= $C$; \\
Initialize $L(i)$ = {gates applied at layer $i$ $\mid$ $0 < i <$ $Depth_C$; }\\
\For{each each layer $i$}{
    \For{each gate $g_{j}$ in $L(i)$ }{
        \For{ each immediate dependent gate $g_{k}$}{
        \If{Are\_commutative($D$, $g_{j}$, $g_{k}$)}{
            \If{No\_Inc\_Depth($D$, $g_{j}$, $g_{k}$)}{
                    Compute WESP of $C^{'}$ under rescheduled gates;
                }         
            }
           
        }
        Select rescheduling of $g_{j}$ with max. WESP;\\
        Update $C^{'}$;
    }
}
Return $C^{'}$;

}
 \end{small}
\end{algorithm}
%\end{wrapfigure}
 % \lipsum

%$M_2$ mapping shown in Figure~\ref{fig:m2wesp}. WESP for $M_2$ is 0.818353 and ESP for both mapping is 0.825585. The PST for m1 and m2 is $54\%$ and $71\%$.
\subsection{Complex gate rescheduling based on WESP for QAOA}
QAOA quantum circuits consist of single-qubit rotations and two-qubit phase gates implemented using ZZ complex gate. Since ZZ gates are commutative~\cite{crooks2018performance,Venturelli_2018}, we exploit the freedom of the ZZ gate placement at higher quantum circuit design level to further reduce the quantum circuit errors.  ZZ gates are reordered without violating the circuit depth and the gate count. 

\begin{figure}[t]
     \centering
 %    \vspace{-0.1in}
     \begin{subfigure}{.4\columnwidth}
         \centering
         \includegraphics[width=1.1in]{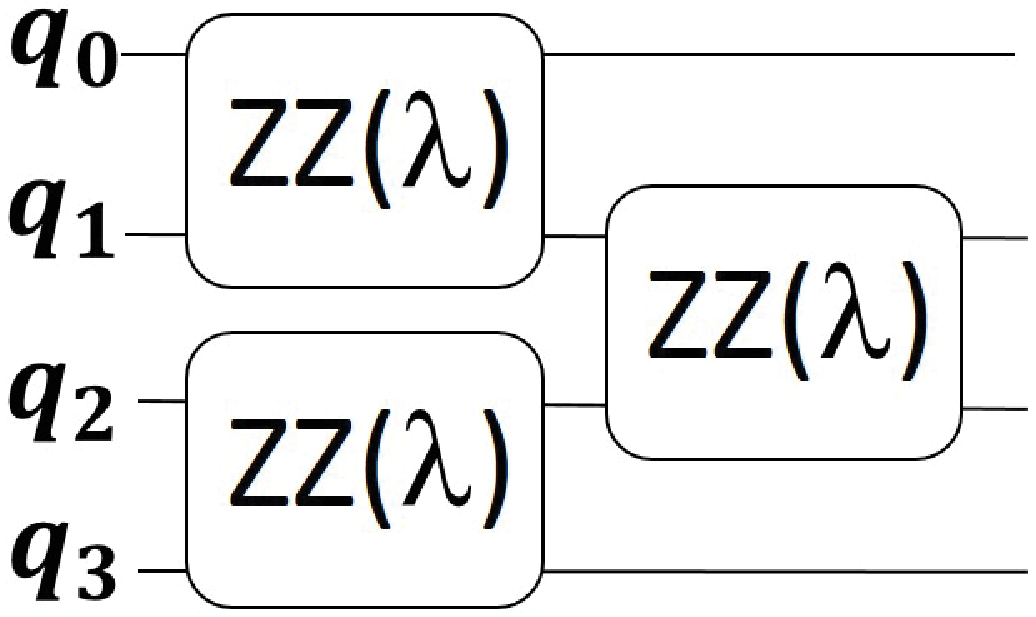}
         \caption{}
         \label{fig:(a)}
     \end{subfigure}
   %  \hfill
     \begin{subfigure}{.4\columnwidth}
         \centering
         \includegraphics[width=1.1in]{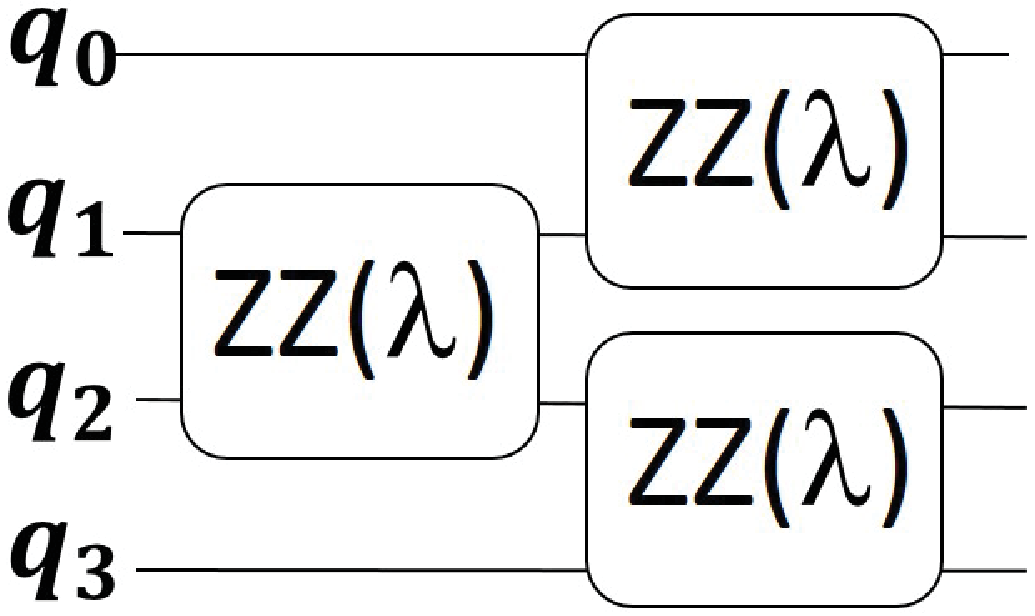}
         \caption{}
         \label{fig:(b)}
     \end{subfigure}
        \vspace{-0.05in}
        \caption{Two quantum sub-circuits with different ZZ gate schedulings.}
        \vspace{-0.05in}
        \label{fig:ZZ}
\end{figure}

\begin{example}
An example of ZZ gate reordering is provided in {Figure}~\ref{fig:ZZ}, in which both quantum sub-circuits share the same gate count and circuit depth but vary in the ZZ gate scheduling.
\end{example}

We propose a look-ahead approach that approximates the error rate of each complex gate in the circuit. The approximation entitles the generation of physical quantum circuit first, using an efficient quantum mapping approach that allocates physical qubits with minimum error rates and minimizes the number of SWAP operations. We scan the physical quantum circuit next to identify the corresponding two-qubit gates used for building each ZZ complex gate. Given the decomposed sub-circuit of each complex gate, we compute the error rate of each complex gate.

We adjust the WESP metric to operate on the intermediate representation of the quantum circuit prior to complex gate decomposition. In the updated WESP, $G$ is the number of ZZ gates, $e_{g_i}$ is the ZZ gate error rate computed as the product of the error rates of its elementary gates, $w_i$ is the ratio of dependent (reachable) ZZ gates, and {\color{black}  $\min(E_{G})$ }is the minimum error rate of the ZZ complex gate in the circuit. The ZZ gates are reordered based on the updated WESP using a similar rescheduling algorithm as the one provided in Algorithm~\ref{alg:alg2}, in which $C$ is updated to be the intermediate representation of the quantum circuit and every pair of quantum gates ($g_i$, $g_j$) corresponds to a pair of ZZ gates in the quantum circuit.
%the mapping approach
%We propose a look ahead-based approach to perform the complex gate-level rescheduling to reduce the Pauli errors by maintaining the depth of the QAOA circuit. We obtain the initial mapped QAOA circuit based on the look-ahead approach by scanning the circuit and look for complex gates that share the same qubits. If the shared qubits are not adjacent, then we introduce SWAP operation to establish the connection to obtained the physically mapped circuit. Finally, we perform complex gate-level rescheduling for the obtained post mapped QAOA based on the proposed WESP metric similar to the Algorithm~\ref{alg:alg2}. 
The look-ahead approach requires scanning the entire physical circuit including all its gates. The time complexity of the look-ahead approach is linear in the number of gates ({\color{black}$O(|V|)$}). The time complexity of our ZZ gate rescheduling algorithm is {\color{black}$O({|V^{'}}|^{2} \cdot (|V^{'}|+|E^{'}|))$} for $D^{'}(V^{'},E^{'})$ complex gate dependency graph, in which $V^{'}$ represents the {\color{black} set} of ZZ complex gates and ${E^{'}}$ is the {\color{black} set} of edges that show the dependency between different ZZ gates. %{\color{black}In our approach, we do not consider the SWAP operations of the circuit while reordering the complex gates. Because it requires updating the circuit mapping every time and the WESP for a potential gate reordering has to be computed, which is a time-consuming process. Instead, we consider reordering the CNOT gates at the elementary gate level.}
%. The complexity of the Algorithm is dominant to the complexity of the look-ahead approach, thus neglecting the complexity of the look-ahead approach, the total time complexity of our approach for complex gate-basedescheduling is $O(V^{2} \cdot (V+E))$.

%\todo{make all D(V,E) as $D(V,E)$. Also, check that all swap operations are written as SWAP operations.}

{\bf We emphasize that our rescheduling approaches, which are applied post-mapping can be integrated with other quantum compilation approaches that exploit the gate commutation rules to reduce the circuit depth and the gate count since our algorithms do not alter the circuit depth nor the gate count. Furthermore, our proposed approaches can be easily applied to quantum circuits executed on different quantum hardware in the presence of significantly variable hardware gate errors. }

\iffalse
The look-ahead approach requires scanning the entire physical circuit including all its gates. The time complexity is linear linear that is the same as scanning the whole dependency graph; the complexity is $O(V+E)$ and the complexity of our greedy heuristic appro
ach is $O(V^{2} \cdot (V+E))$. The complexity of the Algorithm is dominant to the complexity of the look-ahead approach, thus neglecting the complexity of the look-ahead approach, the total time complexity of our approach for complex gate-based rescheduling is $O(V^{2} \cdot (V+E))$.
\fi

\section{Experimental Evaluation}

\subsection{Experimental setup}
%\todo{we need to define the quantum volume since you are showing it in the table and cite its paper too}
%\todo{please add the 3 versions of QAOA to table II}
We show the effectiveness of our proposed rescheduling algorithms based on WESP in improving the output state fidelity of different quantum circuits executed on NISQ computers. We conduct two experiments. In the first experiment, we use PST to evaluate our proposed elementary gate rescheduling approach applied to different quantum circuits, while in the second experiment, we show the impact of our proposed complex ZZ gate rescheduling on the AR of QAOA quantum circuits. 

We use a total of 12 benchmark circuits with different number of qubits and gate count. The \textbf{Bernstein-Vazirani (BV)} algorithm identifies a hidden string, which is encoded in the circuit~\cite{10.1137/S0097539796300921}. The discrete fourier transform is represented using \textbf{Quantum Fourier Transform (QFT)} circuit~\cite{nielsen_chuang_2019}. A shift operation is carried out for a given input Boolean function using \textbf{Hidden Shift (HS)} algorithm~\cite{10.5555/644108.644189}. The \textbf{Grover Search (Grover)} algorithm provides a solution ($x$) of a function $f(x)$ equal to 1. Given a unitary operator, the eigenvalue of the eigenvector is estimated using \textbf{Quantum Phase Estimation (QPE)} algorithm~\cite{nielsen_chuang_2019}. \textbf{Quantum Approximate Optimization Algorithm (QAOA)} solves optimization problems~\cite{farhi2014quantum}. Other reversible quantum circuits such as \textbf{Adder, Toffoli gate, and Decoder} are also used. We obtained QFT and QPE circuits from Qiskit~\cite{Qiskit}, HS and QAOA circuits from Cirq~\cite{quantum_ai_team_and_collaborators_2020_4062499}, adder from~\cite{2004quant.ph.10184C}, decoder from RevLib~\cite{4539430}, and BV, Toffoli, and Grover Search by manual construction. The properties of all the benchmark circuits are provided in Table~\ref{tab:prop} prior to the circuit mapping, which includes the number of qubits (\# Qubits), single-qubit gates (U), CNOT gates (CNOT), and measurement operations (M), the circuit depth (Depth), and the expected output. We use QAOA to solve the Max-Cut problem of three different graphs, which consist of 15 nodes and different number of edges of random weight per node. The QAOA quantum circuits are \textbf{QAOA\_1}, \textbf{QAOA\_2}, and \textbf{QAOA\_3}, which target graphs of 15 nodes with 5, 6, and 7 edges per node, respectively. To construct and test the QAOA circuits, we set $p$ to 1 with default $\gamma$ and $\beta$ parameters provided by Cirq~\cite{quantum_ai_team_and_collaborators_2020_4062499}. Unlike the other quantum circuits listed above, we consider the entire output distribution of QAOA quantum circuits to evaluate their cost function.  
\begin{table}[t]
\centering
\caption{Properties of quantum circuit benchmarks.}
%\vspace{-0.05in}
\label{tab:prop}
%\small
%\resizebox{\textwidth}{!}
{%
\begin{tabular}{|c|c|c|c|c|c|c|}\hline
\multicolumn{1}{|c|}{\multirow{2}{*}{\begin{tabular}[c]{@{}c@{}}Benc-\\hmark\end{tabular}}} & \multicolumn{1}{|c|}{\multirow{2}{*}{\begin{tabular}[c]{@{}c@{}}\#\\Qubits\end{tabular}}}  & \multicolumn{3}{c|}{\#} & \multirow{2}{*}{Depth}&\multirow{2}{*}{\begin{tabular}[c]{@{}l@{}}Expected\\ Output\end{tabular}} \\ \cline{3-5}
 &  & U & CNOT & M & & \\ \hline
BV\_3 & 4 & 9 & 2 & 3 & 6 & 110 \\
Adder & 3 & 30 & 17 & 2 & 28&0100 \\
Grover  & 3 & 24 & 7 & 2 & 22&10 \\ 
Toffoli & 3 & 11 & 6 & 3 & 12&111 \\
QFT\_3 & 3 & 10 & 6 & 3 & 14&000 \\
QFT\_4 & 4 & 26 & 18 & 4 & 28&0000 \\
Decoder & 5 & 28 & 21 & 4 & 29&00100 \\
QPE\_3 & 3 & 19 & 7 & 2 & 17&01 \\
QPE\_4 & 4 & 24 & 14 & 3 & 31&001 \\ 
QAOA\_1 & 15 & 437 & 111 & 15 & 176 & NA \\ 
QAOA\_2 & 15 & 525 & 135 & 15 &214 & NA\\  
QAOA\_3 & 15 & 602 & 156 & 15 &224 & NA \\  \hline 
\end{tabular}
}
%\vspace{-0.1in}
\end{table}

\begin{table}[b]
\centering
\caption{Properties of different IBM quantum computers.}
\label{tab:Res1}
%\resizebox{\textwidth}{!}
{%
\begin{tabular}{|c|c|c|c|c|c|}\hline
\begin{tabular}[c]{@{}l@{}}Quantum Computer\end{tabular} & 
\begin{tabular}[c]{@{}l@{}}\# Qubits\end{tabular} & 
\begin{tabular}[c]{@{}l@{}}QV\end{tabular} \\\hline
IBM Q16 Melbourne & 15 & 8 \\
IBM Q Casablanca & 7 & 32 \\
IBM Q Santiago & 5 & 32 \\
IBM Q Rome & 5 & 32 \\ \hline %%all gates with error 10^-5
\end{tabular}%
}
\end{table}

\begin{figure*}[t]
\centering
     \centering
     \begin{subfigure}{0.5\textwidth}
         \centering
         \includegraphics[width=3.5in]{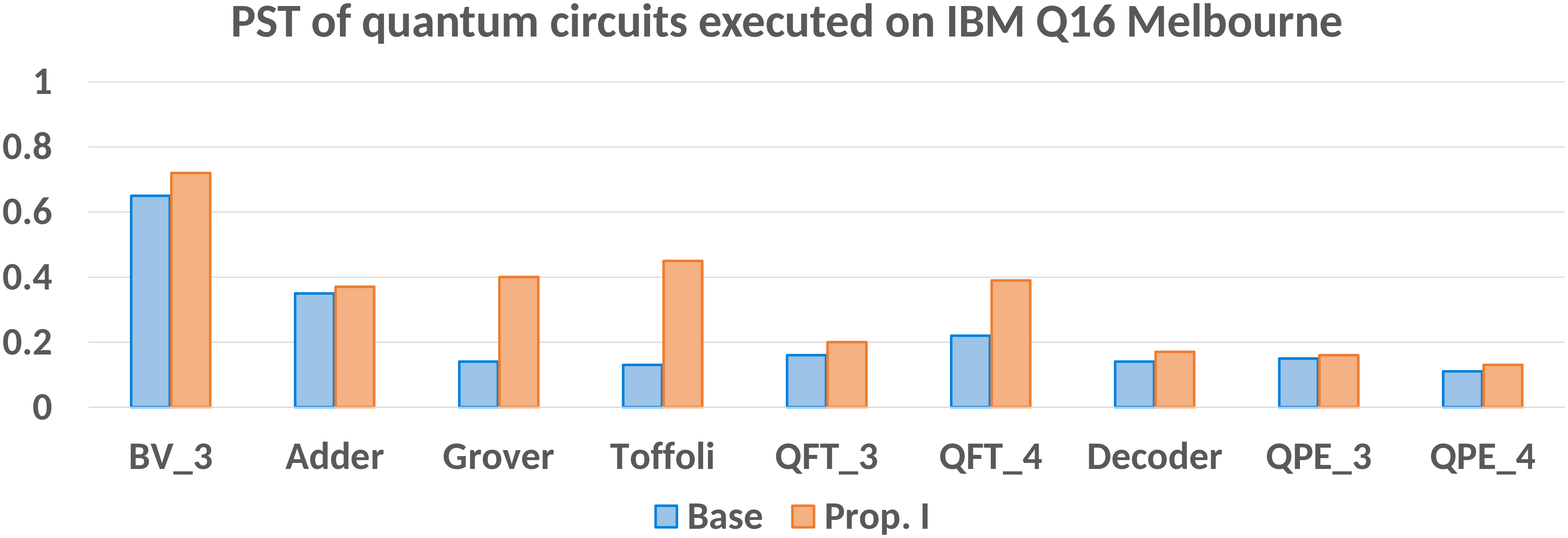}
         \caption{}
         \label{fig:(a)}
     \end{subfigure}\hfill
     \begin{subfigure}{0.5\textwidth}
         \centering
         \includegraphics[width=3.5in]{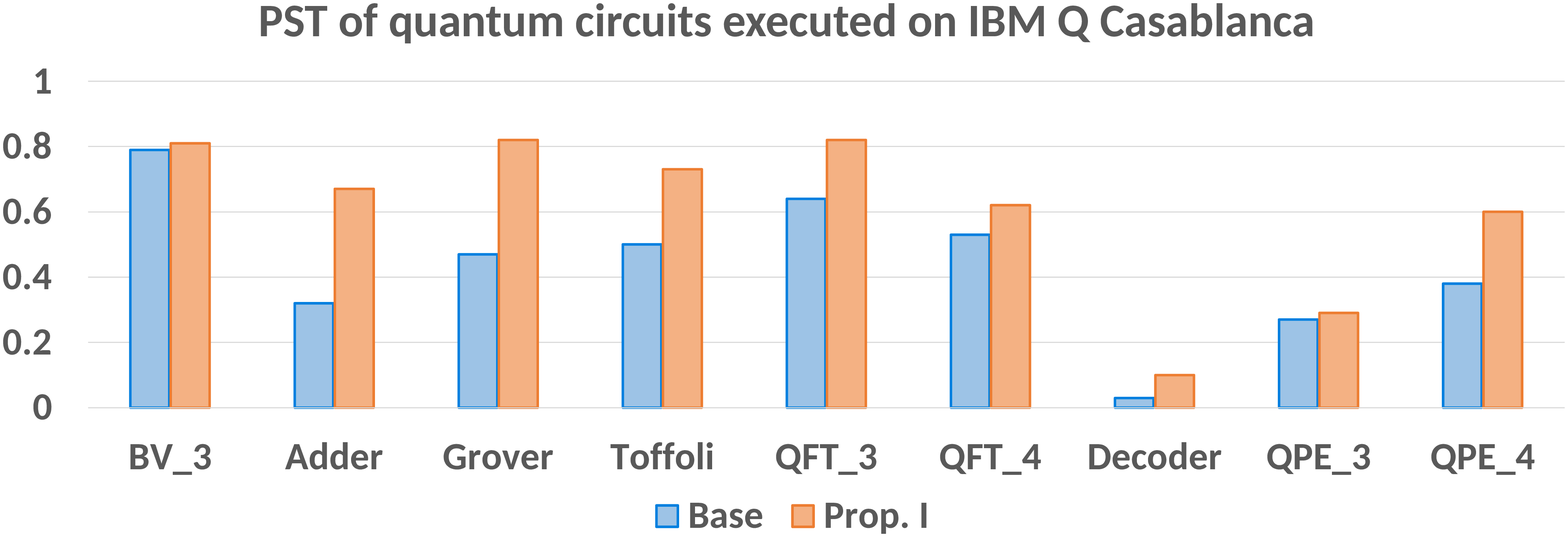}
         \caption{}
         \label{fig:(b)}
     \end{subfigure}\hfill
     \begin{subfigure}{0.5\textwidth}
         \centering
         \includegraphics[width=3.5in]{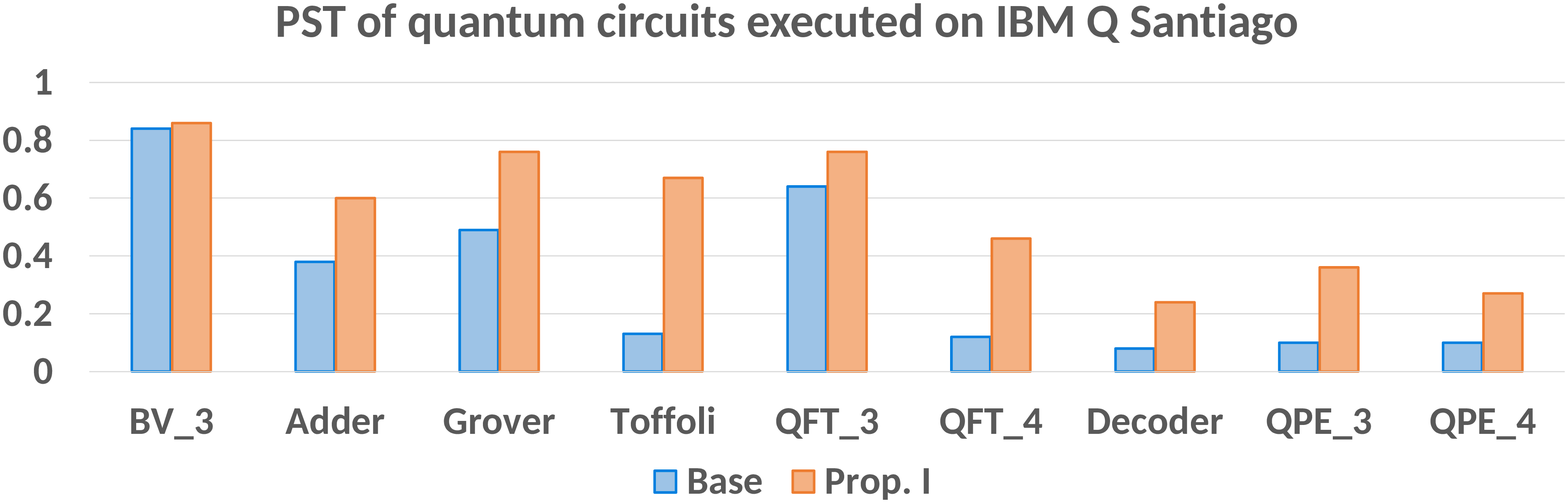}
         \caption{}
         \label{fig:(c)}
     \end{subfigure}\hfill
     \begin{subfigure}{0.5\textwidth}
         \centering
         \includegraphics[width=3.5in]{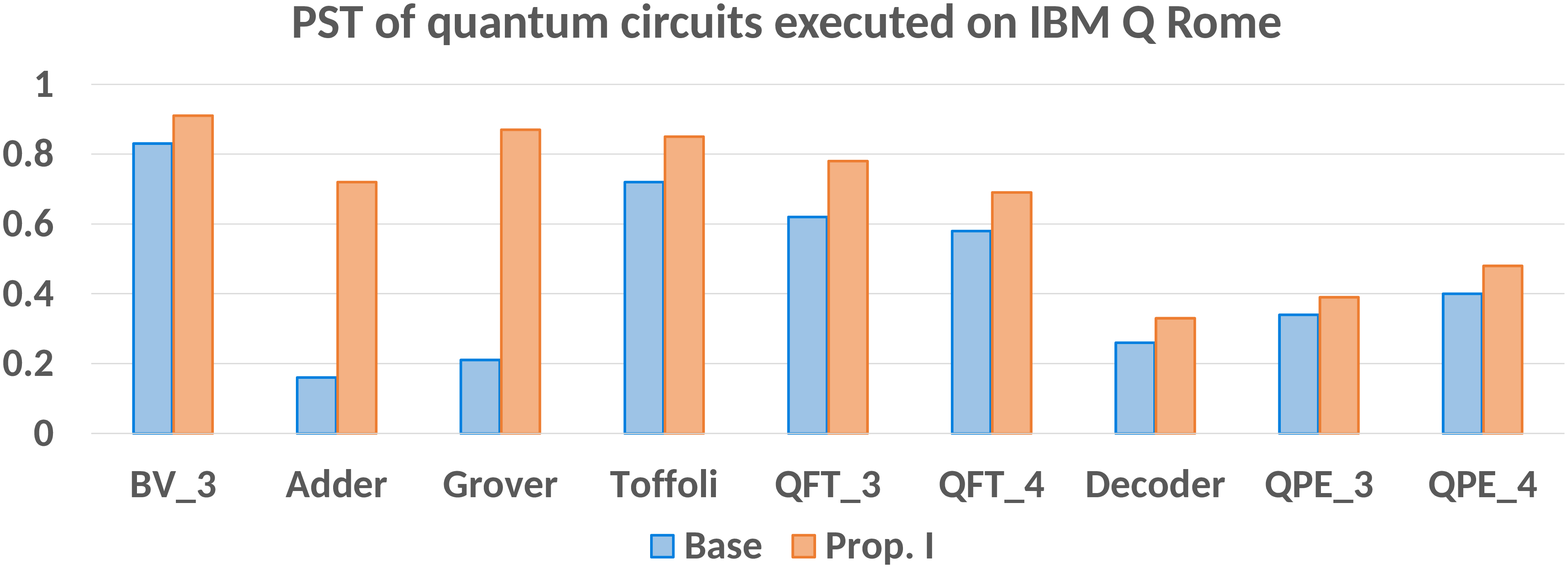}
         \caption{}
         \label{fig:(d)}
     \end{subfigure}
\caption{{Probability of Successful Trails (PST) for different benchmarks executed on various IBMQ architectures}}
\label{fig:PST1}
\end{figure*}

%\todo{fix figure 7(b). It is not aligned with figure 7(a)}

\begin{table*}[h]
\centering

\caption{Properties of physical quantum circuits generated using Prop. I.}
%\vspace{-0.05in}
\label{tab:acc}
%\resizebox{\textwidth}{!}
{%

\begin{tabular}{|@{\hspace*{0.1cm}}c@{\hspace*{0.1cm}}|c|c|c@{\hspace*{0.1cm}}|c@{\hspace*{0.1cm}}|c|c|c@{\hspace*{0.1cm}}|c@{\hspace*{0.1cm}}|c|c|c@{\hspace*{0.1cm}}|c@{\hspace*{0.1cm}}|c|c|c@{\hspace*{0.1cm}}|c@{\hspace*{0.1cm}}|}
\hline
\multicolumn{1}{|@{\hspace*{0.1cm}}c@{\hspace*{0.1cm}}|}{\multirow{3}{*}{\begin{tabular}[c]{@{}c@{}c@{}}Benc-\\hmark\end{tabular}}} & \multicolumn{4}{@{\hspace*{0.1cm}}c@{\hspace*{0.1cm}}|}{Melbourne} & \multicolumn{4}{@{\hspace*{0.1cm}}c@{\hspace*{0.1cm}}|}{Casablanca} & \multicolumn{4}{@{\hspace*{0.1cm}}c@{\hspace*{0.1cm}}@{\hspace*{0.1cm}}|}{Santiago} & \multicolumn{4}{@{\hspace*{0.1cm}}c@{\hspace*{0.1cm}}|}{Rome}\\ \cline{2-17} 
\multicolumn{1}{|@{\hspace*{0.1cm}}c@{\hspace*{0.1cm}}|}{} & \multicolumn{2}{@{\hspace*{0.1cm}}c@{\hspace*{0.1cm}}|}{WESP} & \multicolumn{1}{@{\hspace*{0.1cm}}c@{\hspace*{0.1cm}}|}{\multirow{2}{*}{R}} & \multicolumn{1}{@{\hspace*{0.1cm}}c@{\hspace*{0.1cm}}|}{\multirow{2}{*}{D}} & \multicolumn{2}{@{\hspace*{0.1cm}}c@{\hspace*{0.1cm}}|}{WESP} & \multicolumn{1}{@{\hspace*{0.1cm}}c@{\hspace*{0.1cm}}|}{\multirow{2}{*}{R}} & \multicolumn{1}{@{\hspace*{0.1cm}}c@{\hspace*{0.1cm}}|}{\multirow{2}{*}{D}} & \multicolumn{2}{@{\hspace*{0.1cm}}c@{\hspace*{0.1cm}}|}{WESP} & \multicolumn{1}{@{\hspace*{0.1cm}}c@{\hspace*{0.1cm}}|}{\multirow{2}{*}{R}} & \multicolumn{1}{@{\hspace*{0.1cm}}c@{\hspace*{0.1cm}}|}{\multirow{2}{*}{D}} & \multicolumn{2}{@{\hspace*{0.1cm}}c@{\hspace*{0.1cm}}|}{WESP} & \multicolumn{1}{@{\hspace*{0.1cm}}c@{\hspace*{0.1cm}}|}{\multirow{2}{*}{R}} & \multicolumn{1}{@{\hspace*{0.1cm}}c@{\hspace*{0.1cm}}|}{\multirow{2}{*}{D}}  \\ \cline{2-3} \cline{6-7} \cline{10-11} \cline{14-15} 
\multicolumn{1}{|@{\hspace*{0.1cm}}c@{\hspace*{0.1cm}}|}{} & \multicolumn{1}{@{\hspace*{0.1cm}}l@{\hspace*{0.1cm}}|}{Base} & \multicolumn{1}{l@{\hspace*{0.1cm}}|}{Prop. I} & \multicolumn{1}{|@{\hspace*{0.1cm}}c@{\hspace*{0.1cm}}|}{}& \multicolumn{1}{|@{\hspace*{0.1cm}}c@{\hspace*{0.1cm}}|}{}    & \multicolumn{1}{@{\hspace*{0.1cm}}l@{\hspace*{0.1cm}}|}{Base} & \multicolumn{1}{@{\hspace*{0.1cm}}l@{\hspace*{0.1cm}}|}{Prop. I} & \multicolumn{1}{|@{\hspace*{0.1cm}}c@{\hspace*{0.1cm}}|}{}& \multicolumn{1}{|@{\hspace*{0.1cm}}c@{\hspace*{0.1cm}}|}{}    & \multicolumn{1}{@{\hspace*{0.1cm}}l@{\hspace*{0.1cm}}|}{Base} & \multicolumn{1}{@{\hspace*{0.1cm}}l@{\hspace*{0.1cm}}|}{Prop. I} & \multicolumn{1}{|@{\hspace*{0.1cm}}c@{\hspace*{0.1cm}}|}{} & \multicolumn{1}{|@{\hspace*{0.1cm}}c@{\hspace*{0.1cm}}|}{}    & \multicolumn{1}{@{\hspace*{0.1cm}}l@{\hspace*{0.1cm}}|}{Base} & \multicolumn{1}{@{\hspace*{0.1cm}}l@{\hspace*{0.1cm}}|}{Prop. I} & \multicolumn{1}{|@{\hspace*{0.1cm}}c@{\hspace*{0.1cm}}|}{}& \multicolumn{1}{|@{\hspace*{0.1cm}}c@{\hspace*{0.1cm}}|}{}\\ \hline
BV\_3 & 6.666E-1 & 6.672E-1 & 1 & 34 & 7.064E-1 & 7.071E-1 & 1 & 16 & 8.590E-1 & 8.591E-1  & 1 & 15 & 9.078E-1 & 9.079E-1 & 1 & 15 \\
Adder & 4.243E-1 & 4.945E-1  & 7 & 74 & 5.992E-1  & 6.442E-1  & 4  & 102 & 4.662E-1  & 5.339E-1  & 5 & 69 & 7.330E-1  & 7.743E-1  & 3 & 74 \\
Grover & 3.909E-1  & 3.940E-1  & 3  & 51 & 8.312E-1  & 8.432E-1  & 2  & 56 & 8.052E-1  & 8.218E-1  & 10 & 73 & 8.756E-1  & 8.831E-1  & 1 & 65 \\
Toffoli & 4.246E-1  & 4.292E-1  & 2 & 16 & 7.516E-1  & 7.870E-1  & 2 & 51 & 7.820E-1  & 7.998E-1  & 3 & 45 & 9.006E-1  & 9.107E-1  & 2 & 44 \\
QFT\_3 & 5.940E-1  & 6.278E-1  & 3 & 65 & 7.942E-1 & 8.127E-1  & 2 & 68 & 7.146E-1  & 7.284E-1  & 7  & 71 & 8.681E-1  & 8.833E-1  & 4 & 65 \\
QFT\_4 & 4.784E-1  & 4.789E-1  & 3 & 94 & 6.978E-1  & 7.235E-1 & 2 & 137 & 4.271E-1  & 4.572E-1  & 4 & 150 & 7.604E-1  & 7.800E-1  & 2 & 76 \\
Decoder & 2.886E-2  & 4.869E-2  & 5 & 141 & 2.648E-1  & 3.059E-1  & 3  & 200 & 2.117E-1  & 2.824E-1  & 6 & 215 & 5.079E-1  & 5.705E-1  & 6 & 229 \\
QPE\_3 & 3.430E-1  & 3.438E-1  & 1 & 77 & 7.664E-1  & 7.911E-1 & 2 & 71 & 7.422E-1  & 7.635E-1  & 9 & 92 & 8.374E-1  & 8.377E-1  & 1 & 92 \\
QPE\_4  & 1.528E-1  & 1.705E-1  & 3 & 127 & 6.691E-1  & 7.096E-1  & 3 & 134 & 3.467E-1  & 4.034E-1  & 5 & 117 & 6.842E-1  & 7.232E-1  & 8 & 102 \\

\hline
\end{tabular}
}

\end{table*}

We run our circuits on a variety of quantum architectures with different number of qubits and Quantum Volume (QV), as shown in Table~\ref{tab:Res1}. The QV determines the largest size of random quantum circuits in terms of the number of qubits and the circuit depth, which can be successfully executed on the quantum computer with high fidelity% are measures the error rates and potential of a quantum computer
~\cite{Cross_2019}. Each execution of the quantum circuit consists of 8192 shots/trials. %\todo{please verify. My understanding is that it has to do with the depth and the number of qubits of quantum circuits that can be executed on the hardware. Check the reference and write it nicely}.

 All the benchmark circuits are mapped to the quantum architecture using Qiskit software development kit based on the error rates of the quantum hardware~\cite{Qiskit}. We implement our rescheduling algorithms using Python programming language for easy integration with Qiskit. We run our algorithms on a 2.10 GHz Intel Xeon(R) CPU E5-2620 processor with 62.8 GB memory. %All the benchmark circuits are mapped to the quantum architecture using Qiskit software development kit~\cite{Qiskit} based on the error rates of the quantum hardware.  Our algorithms are implemented using python for easy integration with Qiskit. %, which slows down the execution time. C/C++ can be used to implement the same algorithm to improve the execution time.
 
We execute 150, 153, 121, and 117 pairs of base and reordered circuits on IBM Q Melbourne, IBM Q Santiago, IBM Q Rome, and IBM Q Casablanca, respectively, based on the availability of these devices. The results of all these circuits show that the reordered quantum circuits according to WESP provided in Section V yield a better solution than the base circuits generated by Qiskit. We select a subset of these circuits to show the effectiveness of our approach.

\subsection{Elementary gate rescheduling results}
%\todo{can you replace Orig. with Base in all the figures in this section and the following section. Also, lets change Prop. 1 to Prop. I and Prop. 2 with Prop. II}
In the first experiment, we study the impact of our proposed elementary gate rescheduling on the PST of different benchmark circuits. For each quantum circuit to be executed on a given quantum computer, we first generate the corresponding physical quantum circuit using the highest optimization level of Qiskit based on the calibration data of the quantum hardware. We refer to this approach as a \textit{Base}. Next, we apply our proposed elementary gate rescheduling algorithm based on WESP to the physical quantum circuit to generate a new optimized physical quantum circuit. We refer to this process as \textit{Prop. I}. We run the two physical quantum circuits on the quantum computer and evaluate their PST.  
%We first take the quantum circuit and get the initial mapping for the circuit ($original$). We calculate the WESP of the post-mapped quantum circuit. Then we perform possible reordering based on the proposed approach and obtain the most updated quantum circuit ($proposed$) with WESP greater than or equal to the WESP of the original circuit while maintaining the depth. Later, we execute both ($original$) and ($proposed$) back to back on different quantum architecture and evaluate the PST of both the circuits.
%We execute both quantum circuits back to back before re-calibrating the quantum hardware.

Figure~\ref{fig:PST1} provides a comparison of the PST of quantum circuits generated using Base and Prop. I and executed on different quantum computers.
%The PST for ($original$) and ($proposed$) for different benchmark circuits is shown in Figure~\ref{fig:PST1} for various IBM quantum architectures. 
We also report the WESP of quantum circuits generated using Base and Prop. I, the number of gate reordering using Prop. I (R), and the depth of each physical quantum circuit (D) executed on different quantum computers in Table~\ref{tab:acc}. We emphasize that the depth of the physical quantum circuit prior to and post applying Prop. I is always the same to avoid additional decoherence errors. 

Figure~\ref{fig:PST1} shows that our proposed rescheduling algorithm improves the PST, and thus, the fidelity of the quantum circuit output. We observe significant improvement of PST for some quantum circuits more than others such as Adder, Toffoli, and Grover Search quantum circuits for most of the quantum computers, which implies that the circuit structure can impact the effectiveness of our approach. We also observe that our proposed approach can also perform well for quantum computers with a larger quantum volume, and thus, a better fidelity such as IBM Q Santiago, IBM Q Casablanca, and IBM Q Rome. 
Furthermore, as our approach exploits the variation in gate error rates, which can result in different error rates propagating throughout different circuit paths, it is expected to perform better in the presence of higher gate error standard deviation of the quantum hardware. This is further illustrated in Figure~\ref{fig:PST1}, for which the standard deviation of the two-qubit gate error of IBM Q16 Melbourne, IBM Q Casablanca, IBM Q Santiago, and IBM Q Rome on average are 1.31E-02, 1.86E-02, 1.33E-02, 1.20E-02, respectively, which are considered relatively high. 

%1.22E-2, 2.40E-3, 1.27E-2, and 1.40E-2, respectively. %The low standard deviation of the two-qubit gate error of IBM Q Casablanca explains the limited improvement of our proposed approach.
%\todo{check vedika data again}
%\todo{Comment about why Casablanca is not very good. Is it because of similar back end error rates of CNOT}
%For the results, we infer that when the number of reorders is high, then the PST of the circuit is also improved.

As shown in~\cite{9296804}, the output state fidelity of the quantum circuit can be improved by rescheduling quantum gates as late as possible in the circuit to reduce the usage time of the qubit, and thus, reduce the qubit decoherence errors. 
To ensure that the improvement in PST in Figure~\ref{fig:PST1} is not due to a reduction in the qubit life time, and thus, decoherence errors, we identify the first layer in which each physical qubit is being used in each pair of quantum circuits generated using the Base and the Prop. I approaches. For non-measured qubits we also identify the last layer being used in both physical quantum circuits. For measured qubits, all measurement operations will be at the end of the circuit according to IBM quantum computers. We observe that none of the circuits used in Figure~\ref{fig:PST1} exhibits any reduction in their physical qubit lifetime after running the rescheduling algorithm, which confirms the effectiveness of our approach in reducing the impact of Pauli errors on the output state fidelity of the quantum circuit.
%related to our proposed by our proposed approach is independent than 
%Also, we make sure that the improvement in PST is not because of reducing the lifetime of the qubits; hence the coherence error is not reduced. The ESP will be the same for both circuits, which implies that both the quantum circuits take the same mapping of the quantum architecture and do not adds any additional gates after reordering the gates in the circuit.

The run time of our proposed elementary gate rescheduling algorithm for each benchmark circuit to be executed on various quantum architectures is shown in Table~\ref{tab:exec}. Each
quantum circuit requires only a small fraction of a second for gate rescheduling time.

%\todo{expand the discussion here as discussed (impact on coherence). Also, you should indicate somewhere in the setting that you develop your code using python for easy integration with qiskit which slow down the execution time. Similar algorithm implemented in C/C++ will be faster}

%\todo{update figure 7. The name of each machine should be consistent to the name in table 1. Check table 1 which I updated}

\begin{table}
\centering
\caption{Run time of Prop. I algorithm applied to different quantum circuits mapped to different IBM Q computers}
%\vspace{-0.05in}
\label{tab:exec}
\begin{tabular}{|l|r|r|r|r|}
\hline
\multicolumn{1}{|c|}{\multirow{2}{*}{\begin{tabular}[c]{@{}c@{}c@{}}Benc-\end{tabular}}} & \multicolumn{4}{c|}{Prop. I Execution Time (msec)} \\ \cline{2-5}
\multicolumn{1}{|c|}{hmark} & \multicolumn{1}{l|}{\begin{tabular}[c]{@{}c@{}}Melb-\\ourne\end{tabular}} & \multicolumn{1}{l|}{\begin{tabular}[c]{@{}c@{}}Casa-\\blanca\end{tabular}} & \multicolumn{1}{l|}{\begin{tabular}[c]{@{}c@{}}San-\\tiago\end{tabular}}& \multicolumn{1}{l|}{Rome}  \\
\hline
BV\_3& 0.54  & 0.17   & 0.17 & 0.23         \\
Adder& 1.25  & 2.17   & 1.37 & 1.52         \\
Grover  & 1.14  & 1.27   & 1.52 & 1.34         \\
Toffoli & 0.20  & 0.53   & 0.85 & 0.90         \\
QFT\_3  & 1.12  & 1.38   & 1.49 & 1.26         \\
QFT\_4  & 1.47  & 2.59   & 2.92 & 1.57         \\
Decoder & 2.52  & 3.27   & 3.11 & 3.28         \\
QPE\_3  & 1.57  & 1.51   & 1.93 & 1.72         \\
QPE\_4  & 2.51  & 1.86   & 2.32 & 1.89  \\    
\hline
\end{tabular}
\end{table}

\subsection{Complex gate rescheduling results}

%\todo{Update: please check my email from Saturday carefully. you don't start the y-axis from zero. Start from whatever I show in that histogram that I sent you}.
\begin{figure}
     \centering
  \begin{subfigure}[b]{0.5\textwidth}
         %\centering
         \includegraphics[width=3.3 in]{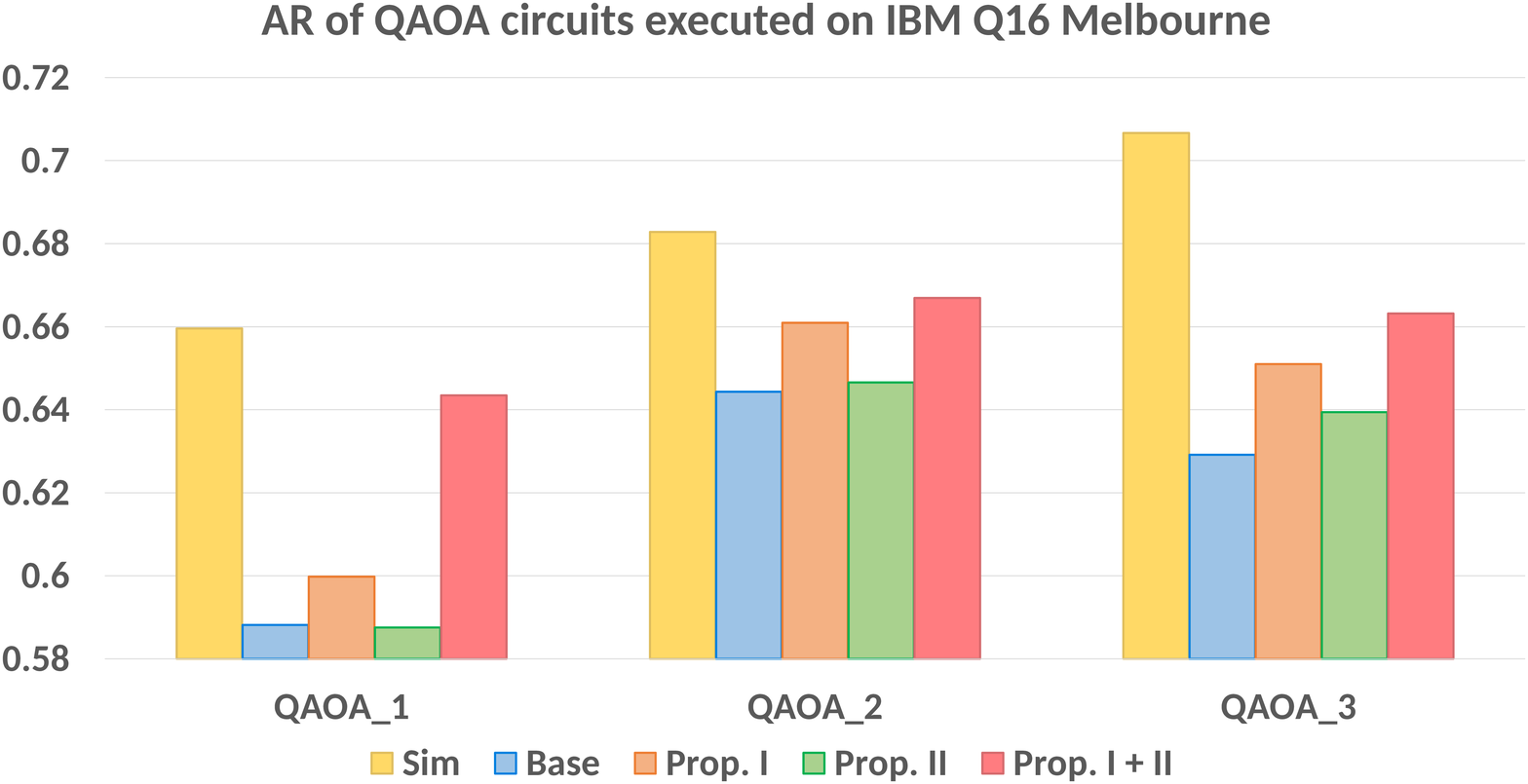}
         \caption{}
         \label{fig:a1}
     \end{subfigure} \vspace{1mm}
     %\hfill
     \begin{subfigure}[b]{0.5\textwidth}
         \centering
         \includegraphics[width=3.3 in]{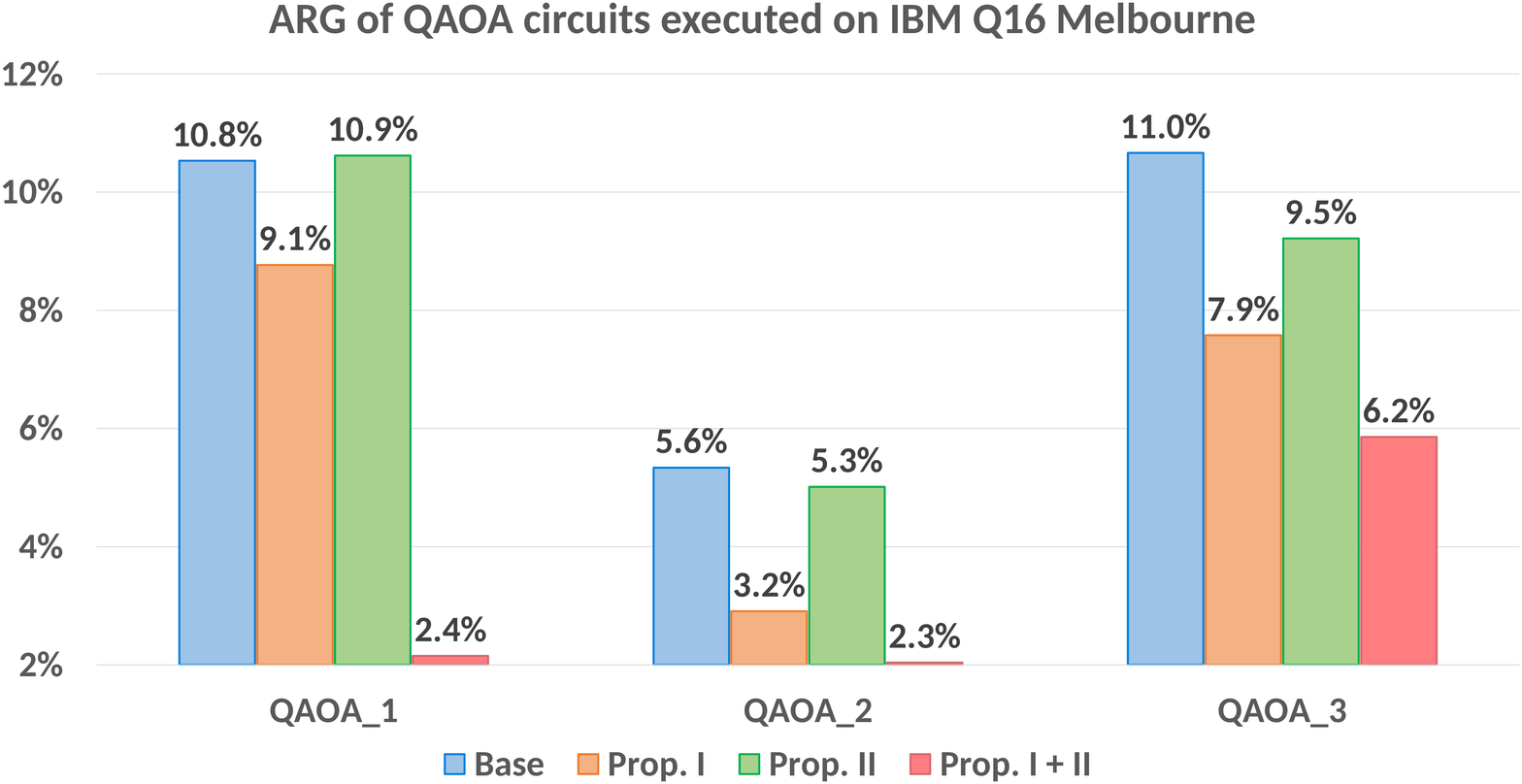}
         \caption{}
         \label{fig:a2}
     \end{subfigure}%\hspace{1mm}
        \caption{(a) The approximation ratio, and (b) the corresponding approximation ratio gap for different QAOA circuits.}
       %\vspace{-0.05in}
        \label{fig:approxqaoa}
\end{figure}

%\todo{you need to define the approximation ratio gap and cite its paper too}
%\todo{you need to replace QAOA\_15q\_7edges, QAOA\_15q\_6edges, and QAOA\_15q\_5edge with QAOA\_1, QAOA\_2, and QAOA\_3 in all the figures}
%\todo{since we define ARO and AR, we should use them for the title of the figure8 ("AR of QAOA circuits executed on IBM Q16 Melbourne" and "ARG of QAOA circuits executed on IBM Q16 Melbourne"}
In the second experiment, we show the effectiveness of our proposed ZZ complex gate rescheduling algorithm with respect to AR, and the corresponding ARG of different QAOA quantum circuits, namely QAOA\_1, QAOA\_2, and QAOA\_3, executed on IBM Q16 Melbourne quantum computer. We refer to the complex gate rescheduling algorithm as \textit{Prop. II}. We also study the impact of applying Prop. II followed by Prop. I approach to generate the physical QAOA quantum circuit, which implies the use of ZZ complex gate rescheduling algorithm first followed by elementary gate rescheduling algorithm after complex gate decomposition based on the commutation rules in Figure~\ref{fig:three graphs}. This process is referred to as \textit{Prop. I+II}.

%To show the impact of ZZ gate rescheduling on the approximate ratio, and thus the corresponding approximate ratio gap (ARG) of QAOA quantum circuits, we generate physical quantum circuits using Base, Prop. I, Prop. II, and the combination of both Prop. I and Prop. II, referred to as \emph{Prop. I+II}.
%We use 15 qubits QAOA circuit for 5edges, 6edges, and 7edges different graph for each qubit to evaluate the performance of three proposed approaches based on approximation ratio and approximation ratio gap (ARG).

%The elementary gate-level reordering approach is referred to as \emph{Prop. I}, the complex gate-level reordering is referred to as \emph{Prop. II} and the complex gate level reordering followed by elementary gate-level is referred to as \emph{Prop. I+II}. 
Figure~\ref{fig:approxqaoa}(a) and (b) provide a comparison of the AR, and the corresponding ARG, respectively, of different QAOA circuits generated using Base, Prop. I, Prop. II, and Prop. I+II.
%The approximation ratio for simulation and execution and approximation ratio gap (ARG) for the execution of different proposed approaches is shown in Figure~\ref{fig:approxqaoa} (a) and (b), respectively.
%We execute the circuits on IBM Q16 Melbourne architecture to extract the results. 
%\todo{Update: can we do one more run under the new condition that you add for QAOA\_1 to improve the complex gate results as you did for QAOA\_2}
The depth of physical QAOA\_1, QAOA\_2, and QAOA\_3 quantum circuits are 315, 375, and 412, respectively, for all the proposed rescheduling approaches. 
 The range of the number of reordering for QAOA\_1, QAOA\_2, and QAOA\_3 quantum circuits are 3-35, 5-39, and 7-53, respectively. %\todo{check the number of reordering of QAOA\_2 since you updated its results} % \todo{update: write the range for each circuit individually} 
%The range of the number of reordering for QAOA\_3, QAOA\_2, and QAOA\_1 quantum circuits are \todo{update: write the range for each circuit individually} 7-53, 6-40, and 3-35, respectively. 
%all the circuits is 3 - 40. 

The WESP computed at the complex and the elementary levels of the base and reordered QAOA quantum circuits and their corresponding ESP are provided in Table~\ref{tab:wesp_qaoa}. 

Our results show that while ZZ gates rescheduling algorithm improves the AR of the QAOA circuits, and thus, reduces the ARG, applying elementary gate rescheduling can provide a better output state fidelity. Our results also show that using both rescheduling algorithms at the complex and the elementary gate-level can deliver the best reduction in the ARG, and therefore, the best output state fidelity. %\textcolor{blue}{The WESP of QAOA circuits are shown in Table~\ref{tab:wesp_qaoa}}
Thus, as both decoherence and gate errors contribute to the quantum circuit noise, by incorporating our gate rescheduling approaches that push very noisy gates to as later circuit layers as possible we can improve the output state fidelity even for QAOA circuits with large depth. 

\begin{table}[]
\centering
{
\caption{{WESP of QAOA quantum circuits generated using the Base, Prop.I, Prop. II, and Prop. I+II and their corresponding ESP value.}}
%\vspace{-0.05in}
\label{tab:wesp_qaoa}
\begin{tabular}{|@{\hspace*{0.08cm}}l@{\hspace*{0.1cm}}|@{\hspace*{0.008cm}}c@{\hspace*{0.08cm}}|@{\hspace*{0.008cm}}c@{\hspace*{0.08cm}}|@{\hspace*{0.01cm}}c@{\hspace*{0.08cm}}|@{\hspace*{0.008cm}}c@{\hspace*{0.08cm}}|@{\hspace*{0.009cm}}c@{\hspace*{0.08cm}}|@{\hspace*{0.01cm}}c@{\hspace*{0.08cm}}|}
\hline
\multicolumn{1}{|c|}{\multirow{2}{*}{\begin{tabular}[c]{@{}c@{}c@{}}Benc-\\hmark\end{tabular}}} & \multicolumn{1}{c|}{\multirow{2}{*}{ESP}} & \multicolumn{2}{l|}{WESP (complex)} & \multicolumn{3}{c|}{WESP (elementary)} \\
\cline{3-7}
\multicolumn{1}{|c|}{hmark} &\multicolumn{1}{c|}{} & \multicolumn{1}{l|}{Base} & \multicolumn{1}{@{\hspace*{0.1cm}}l@{\hspace*{-0.8cm}}|}{Prop. II}& \multicolumn{1}{l|}{Base} & \multicolumn{1}{@{\hspace*{0.02cm}}l@{\hspace*{0.2cm}}|}{Prop. I} & \multicolumn{1}{@{\hspace*{0.02cm}}l@{\hspace*{0.02cm}}|}{Prop. I+II}  \\

\hline
QAOA\_1  & 1.03E-04  & 5.01E-05 & 6.32E-05        & 1.69E-05  & 5.09E-05  & 7.19E-05      \\
QAOA\_2 & 6.81E-07  & 2.49E-07  & 4.58E-07        & 5.61E-08  & 2.31E-07  & 5.36E-07      \\
QAOA\_3 & 9.05E-10 & 3.78E-10  & 5.63E-10       & 1.66E-11  & 4.98E-10  & 7.92E-10  \\
\hline
\end{tabular}
}
\end{table}

The run time of our proposed rescheduling approaches for QAOA\_1, QAOA\_2, and QAOA\_3 quantum circuits to be executed on IBM Q16 Melbourne quantum computer is shown in Table~\ref{tab:qaoa}. Our proposed approaches consume only a small fraction of a second despite the large circuit depth and the number of qubits.

\begin{table}
\centering
\caption{Run time of Prop. I, Prop. II, and Prop. I+II algorithms for different QAOA circuits mapped to IBM Q16 Melbourne.}
\label{tab:qaoa}
\begin{tabular}{|l|r|r|r|}
\hline
\multicolumn{1}{|c|}{\multirow{2}{*}{\begin{tabular}[c]{@{}c@{}c@{}}Benc-\\hmark\end{tabular}}} & \multicolumn{3}{c|}{Execution Time (msec)}   \\
\cline{2-4}
\multicolumn{1}{|c|}{hmark}       & \multicolumn{1}{l|}{Prop. I} & \multicolumn{1}{l|}{Prop. II} & \multicolumn{1}{l|}{Prop. I+II}  \\
\hline
QAOA\_1 & 3.15 & 3.01 & 6.17    \\
QAOA\_2 & 3.22 & 3.11 & 6.35    \\
QAOA\_3 & 3.53 & 3.37 & 7.26   \\
\hline
\end{tabular}
\end{table}
\section{Conclusion}

In this paper, we propose gate rescheduling algorithms based on the gate error propagation paths in the quantum circuits. Given the variation in the error rates of NISQ computers, we show that the location of the quantum gate can significantly affect the output state fidelity of the quantum circuit. Our proposed approaches can be easily integrated with other quantum compilation approaches, which ensure that the quantum circuit mapping process maximizes ESP. We also anticipate further improvements in the output state fidelity when applying our approaches with other compilation approaches that target decoherence and correlated errors. Our future work will further investigate the quantum circuit structures that benefit the most from our proposed gate rescheduling algorithms.

\iffalse
In this paper, we propose gate rescheduling algorithms based on the impact of the gate error propagation paths on the output state fidelity of quantum circuits executed on quantum computers with variable error rates. Our proposed approaches can be easily integrated with other quantum compilation approaches that carefully map quantum circuits to the target hardware to maximize ESP. We anticipate further improvements in the output state fidelity when applying our approaches with other compilation approaches that target decoherence errors and correlated errors. Our future work will further investigate the quantum circuit structures that benefit the most from our proposed gate rescheduling algorithms.

\fi

\section{Acknowledgement}
We thank IBM Quantum for the access to the IBM processors through the IBM Quantum Researchers Program.

\bibliographystyle{IEEEtran}
\bibliography{References}
\end{document}